\documentclass[aps,prl,secnumarabic, nobibnotes, twocolumn,superscriptaddress]{revtex4-1}

\usepackage{amsfonts}
\usepackage{mathrsfs}
\usepackage{amsmath}
\usepackage{color}
\usepackage{natbib}
\usepackage{graphicx}
\usepackage{bm}
\usepackage{amssymb}
\usepackage{xspace}
\usepackage{epstopdf}
\usepackage{dcolumn}
\usepackage{longtable}
\usepackage{multirow}
\usepackage[colorlinks=true, letterpaper=true, pdfstartview=FitV, linkcolor=blue, citecolor=blue, urlcolor=blue]{hyperref}

\makeatletter

\newcommand{\Rmnum}[1]{\expandafter\@slowromancap\romannumeral #1@}

\makeatother

\begin{document}

\title{
Abundant surface-semimetal phases in three-dimensional obstructed atomic insulators}

\author{Xianyong Ding}
\altaffiliation{X. Y. D. and X. J. contributed equally to this work.}
\affiliation{Institute for Structure and Function $\&$ Department of Physics, Chongqing University, Chongqing 400044, P. R. China}
\affiliation{Center of Quantum materials and devices, Chongqing University, Chongqing 400044, P. R. China}

\author{Xin Jin}
\altaffiliation{X. Y. D. and X. J. contributed equally to this work.}
\affiliation{College of Physics and Electronic Engineering, Chongqing Normal University, Chongqing 400044, China}
\affiliation{College of Materials Science and Engineering, Chongqing University, Chongqing 400044, China}

\author{Zhuo Chen}
\affiliation{Institute for Structure and Function $\&$ Department of Physics, Chongqing University, Chongqing 400044, P. R. China}
\affiliation{Center of Quantum materials and devices, Chongqing University, Chongqing 400044, P. R. China}

\author{Xuewei Lv}
\affiliation{College of Materials Science and Engineering, Chongqing University, Chongqing 400044, China}

\author{Da-Shuai Ma}
\email{madason.xin@gmail.com}
\affiliation{Institute for Structure and Function $\&$ Department of Physics, Chongqing University, Chongqing 400044, P. R. China}
\affiliation{Center of Quantum materials and devices, Chongqing University, Chongqing 400044, P. R. China}

\author{Xiaozhi Wu}
\email{xiaozhiwu@cqu.edu.cn}
\affiliation{Institute for Structure and Function $\&$ Department of Physics, Chongqing University, Chongqing 400044, P. R. China}

\author{Rui Wang}
\affiliation{Institute for Structure and Function $\&$ Department of Physics, Chongqing University, Chongqing 400044, P. R. China}
\affiliation{Center of Quantum materials and devices, Chongqing University, Chongqing 400044, P. R. China}

\begin{abstract}
Three-dimensional (3D) obstructed atomic insulators (OAIs) are characterized by the appearance of floating surface states (FSSs) at specific surfaces. 
Benefiting from this feature, our study here shows the presence of abundant surface-semimetal phases in 3D OAI.
The symmetries of obstructed Wannier charge centers ensure the degeneracy of such FSSs at high symmetry points or invariant lines in the surface Brillouin zone. 
Utilizing topological quantum chemistry theory, we identify a carbon allotrope with body-centered tetragonal structure, named as bct-C$_{20}$, as an ideal candidate for realizing different kinds of surface-semimetal phases. For the (001) surface of bct-C$_{20}$, 
there are four in-gap FSSs, and these four FSSs form two kinds of surface Dirac cones, \textit{i.e.}, topological Dirac cones with linear dispersion and symmetry-enforced quadratic Dirac cones. 
The band topology of surface Dirac cone is captured by the effective surface Hamiltonian and the emergence of hinge states.
Moreover, the existence of surface-nodal-line state is also discussed.
This work reports a new approach to obtain $d$-dimensional semimetal phases from the surface states of $(d+1)$-dimensional systems, which is of great significance for the studies in revealing topological states and their practical applications in high-dimensional crystals.

\end{abstract}



\maketitle

\textit{Introduction. ---}
Two-dimensional  topological semimetal states enabled by crystalline symmetry are currently one of the most active research areas in condensed matter physics ~\cite{neto2009electronic, palumbo2015two, young2015dirac, wang2015strong,  feng2017experimental, gao2018epitaxial, jeon2019ferromagnetic, feng2019discovery, niu2019mixed, jin2020two}. Such semimetals are characterized by  band topology protected band nodes with distinct energy dispersion, degree of degeneracy, and the dimension of the degeneracy manifold \cite{wan2011topological, chiu2014classification, bansil2016colloquium, burkov2016topological, PhysRevB.101.195111}.
Systems with topologically non-trivial band nodes are proposed to possess many exotic physical properties such as ultrahigh carrier mobility \cite{xiao2010berry, hosur2013recent, PhysRevB.93.165127, hu2019transport}, half-integer quantum Hall effects \cite{novoselov2005two, zhang2005experimental, novoselov2007room, zou2022half}, large diamagnetism and electromagnetic duality \cite{Singh_2022, fujiyama2022large}.
Thus, such eyetems are regarded as potential candidates for the next-generation quantum information technology revolution \cite{RevModPhys.81.109, ali2014large, lai2018anisotropic, wang2019direct, culcer2020transport}. Nevertheless, the experimental research on 2D topological states lags behind the theoretical one seriously~\cite{novoselov2005two,zhang2005experimental, doi:10.1126/science.1137201, young2015dirac, ren2016topological, wang2019two, ma2019experimental}. 
The limited number of experimentally synthesized 2D topological materials, the non-negligible spin-orbit coupling, and the inevitable coupling to the substrate are the main reasons for this status quo. 
Therefore, achieving high-quality 2D topological semimetal states remains an ongoing question that needs to be addressed urgently.

Most recently, with the developing of topological quantum chemistry (TQC) \cite{bradlyn2017topological, elcoro2021magnetic}, symmetry indicator\cite{po2017symmetry, kruthoff2017topological, watanabe2018structure, song2018quantitative}, real space invariant (RSI) \cite{song2020twisted, xu2021three} and related theories \cite{peng2022topological, bouhon2021topological}, the rapid developments in revealing the link between the emergence of quasi-particles and symmetry constraints are witnessed by both physics and material science communities. 
These theories have undoubtedly made a major breakthrough in revealing the band degeneracy in the bulk states of 2D systems.
In addition to the bulk state, we realize that it is feasible to realize $d$-dimensional semimetal states in the surface states of a $(d+1)$-dimensional obstructed atomic insulators (OAIs) \cite{li2022atomic, wang2022two} or unconventional materials \cite{gao2022unconventional}.
In particular, the OAI is featured by the mismatch between Wannier charge centers and atom-occupied Wyckoff positions (WPs) \cite{doi:10.1126/science.aaz7650, xu2021filling, xu2021three}.
Thus, this kind of non-trivial phase can be effectively diagnosed as an insulator whose band representations (BRs) can be expressed as a sum of elementary BRs (EBRs) with a $necessery$ one induced by $orbitals$ locating at empty position $W$.
$W$ is named as obstructed Wannier charge centers (OWCCs) which can be easily identified by RSI theory. Presently, various novel properties in OAIs have been proposed in both theoretical and experimental research, such as electrodes \cite{PhysRevB.103.205133},  catalysis \cite{xu2021three}, large shift current \cite{zhang2022large}, superconductivity \cite{PhysRevLett.124.247001}, and beyond~\cite{PhysRevResearch.2.033224,PhysRevB.106.155144,PhysRevB.105.165135,GAO2022598,PhysRevLett.124.063901,kim2023bismuth}. Most importantly, with cleavage terminations cutting through OWCCs, floating surface states (FSSs) could appear within the bulk gap~\cite{PhysRevResearch.2.033224,PhysRevB.105.165135,GAO2022598}. However, it is still unclear whether the 2D quasi-particles could emerge in the surface states of 3D OAIs, and if so, which system would be a candidate.

Here, we put forward a strategy to explore abundant 2D quasi-particles on surfaces of a 3D OAI. Due to the map from the 3D space group to 2D layer group, we first elucidate that the 2D quasi-particles originate from a surface lattice formed by $orbitals$ at OWCCs, whose symmetry guarantees that the FSSs degenerate at high symmetry points or invariant lines in the surface Brillouin zone (BZ). We further show by first-principles calculations that a novel carbon allotrope, \textit{i.e.}, bct-C$_{20}$, can achieve our proposal. The energetic, mechanical, dynamical, and thermodynamic stability of bct-C$_{20}$ are all verified.
We reveal that the electronic band structure of  bct-C$_{20}$ has a wide band gap ($>$1~eV), which was proved to belong to the OAI phase. Four in-gap FSSs appear at the (001) surface.  
One can simultaneously observe a topological Dirac cone with linear dispersion and symmetry-enforced quadratic Dirac cone formed by these four FSSs. 
The symmetry protection mechanism of these surface Dirac cones is detailed discussed. 
The topology feature of the topological surface Dirac cone is depicted by the appearance of a hinge state and effective surface Hamiltonian.

\begin{figure}[t]
	\includegraphics[width=\linewidth]{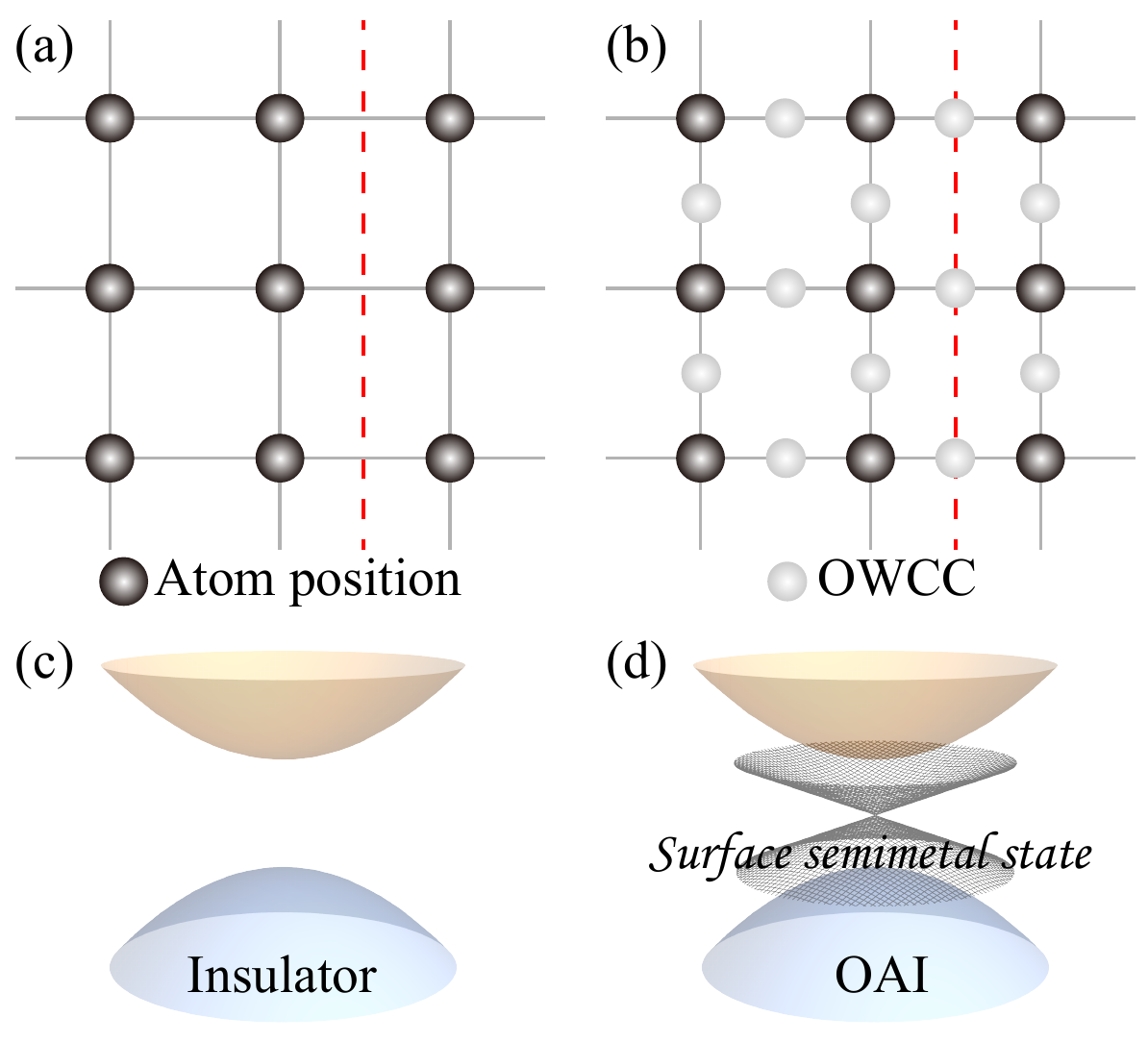}
	\caption{
 (a)-(b) The sketch plot of an atomic insulator (a) and OAI (b). The atom-occupied position and the OWCC are represented by the spheres in black and white, respectively. (c) There is no surface state in the bulk gap when the atomic insulator is cleaved. (d) When the OWCC is exposed on the surface, the 2D semimetal state emerges potentially. The cleavage terminations discussed in (c) and (d) are represented by the red dashed line in plots (a) and (b).
 }
\label{fig1}
\end{figure}

\begin{figure}[tb]
\includegraphics[width=\linewidth]{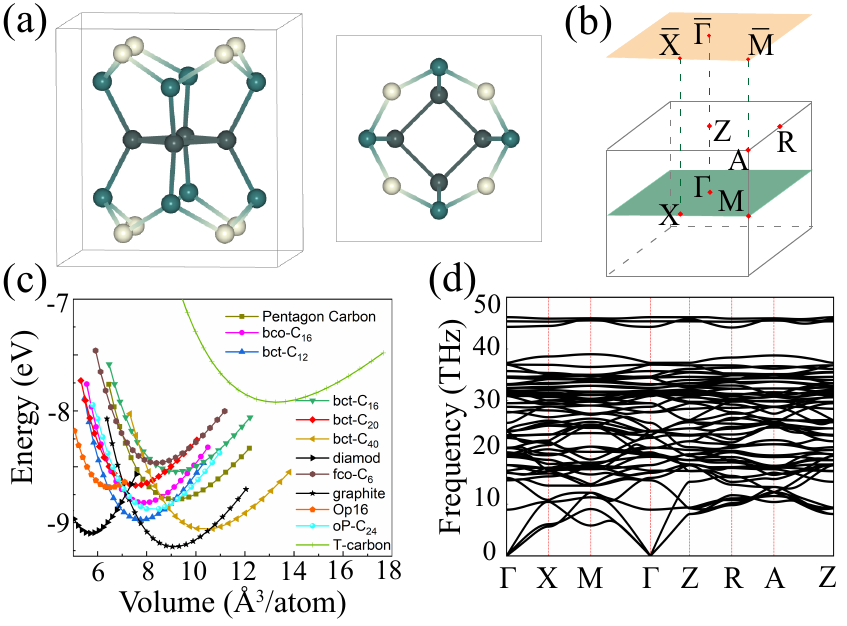}
\caption{
Crystal structure of three dimension bct-C$_{20}$.
(a) Bird and side views of the conventional unit cell for bct-C$_{20}$ with space group of $P4/mmm$ (No. 123). In this plot, different colors denote different Wyckoff positions.
(b) The BZ and the corresponding (001) surface BZ of bct-C$_{20}$.
(c) Energy versus volume of bct-C$_{20}$ and other carbon analogues, \textit{i.e.}, cubic diamond, T-carbon, graphite, bct-C$_{12}$, bct-C$_{16}$, bct-C$_{20}$, bco-C$_{16}$, fco-C$_{6}$, bct-C$_{40}$, oP-C$_{24}$, Pentagon Carbon and Op16 are also shows for comparison.
(d) The phonon spectrum of bct-C$_{20}$.}
\label{fig2}
\end{figure}

\textit{Argument to reach 2D topological semimetal phases. ---}
We start with an intuitive argument to elucidate the physical mechanism of how to reach the 2D semimetal phase in the surface states of 3D OAIs. Here, we ignored the spin-orbital couplings, a similar argument could be extended to the spinful case.
As schematically shown in Figs.~\ref{fig1}(a)-\ref{fig1}(b), in sharp contrast to atomic insulators, there are some $orbitals$ pinned at the OWCCs in OAIs.
When cleaved with these orbitals exposed on a specific surface, there would be FSSs as shown in Fig.~\ref{fig1}(d).
Physically, if the total number of $orbitals$ exposed on the surface is odd, the emerging FSSs would be metallic, which is referred to as filling anomaly \cite{benalcazar2019quantization}.
As a comparison, when the number is even, each valence band is fully occupied.
Then, there are two potential cases where we can find 2D quasi-particles.
\textbf{$Case$-I}: Two FSSs degenerate at high symmetry point $\overline{\mathrm{K}}_s$ in the surface BZ and form a 2D irreducible representation of the little group at $\overline{\mathrm{K}}_s$.
\textbf{$Case$-II}: With the additional crystalline symmetry (\textit{e.g.}, mirror symmetry that normals to the surface) survived in the surface, there naturally would be symmetry (mirror) protected 2D semimetal states when two FSSs with opposite symmetry eigenvalues cross on the symmetry (mirror) invariant line in surface BZ.
The 2D nodal line state formed by surface states of OAI and constrained by the non-symmorphic symmetries is briefly discussed in \textit{Discussion and Summary}. In this work, both of the two cases are verified in  bct-C$_{20}$.

\textit{The Structure, Stability and Electronic Properties of bct-C$_{20}$. ---}
The crystalline structure of bct-C$_{20}$ is shown in Fig.~\ref{fig2}(a).
This system comprises 20 atoms arranged in a body-centered tetragonal structure with space group $P4/mmm$ (No.123).
The optimized lattice constants are $a = b = 5.202$ {\AA} and c = 5.578 {\AA}.
The carbon atoms in this system are distributed over three inequivalent WPs, \textit{i.e.}, $4o$ $(1/2, x, 1/2)$, $8t$ $(x, 1/2, z)$, and $8r$ $(x, x, z)$.
As shown in Fig.~\ref{fig2}(a), the unit cell of bct-C$_{20}$ is composed of four- and five-sided rings, and the adjacent cells are connected by hexagonal rings.

Here, we verify the energetic, mechanical, dynamical, and thermodynamic stability of bct-C$_{20}$.
Comprehensive and detailed data supporting stability are available in the supplementary material (SM) \cite{SM}.
As shown in Fig.~\ref{fig2}(c), the total energies of different carbon analogs are obtained.
The energy minimal $-8.67$~eV/atom of bct-C$_{20}$ is less than that of T-carbon\cite{T-carbon}, bct-C$_{16}$\cite{bct-C16} and fco-C$_{6}$ \cite{Fco-C6}.
The cohesive energy listed in TABLE S1~\cite{SM} also shows that bct-C$_{20}$ has lower cohesive energy than some reported carbon analogues, \textit{e.g.}, T-carbon\cite{T-carbon}, bct-C$_{16}$\cite{bct-C16} and fco-C$_{6}$ \cite{Fco-C6}, which indicates the energetic stability of bct-C$_{20}$. 
The mechanical stability is confirmed by the independent elastic constants that satisfy the Born mechanical stability for the orthogonal crystal \cite{PhysRevB.90.224104,SM}.
The dynamical stability of bct-C$_{20}$ can be reflected by the phonon spectrum.
As shown in Fig.~\ref{fig2}(d), the phonon branches of bct-C$_{20}$ have no imaginary frequency in the whole BZ.
The AIMD simulations with 10~ps timescale at $T=300$ K demonstrate the thermodynamical stability of bct-C$_{20}$.

\begin{figure}[!htb]
\includegraphics[width=\linewidth]{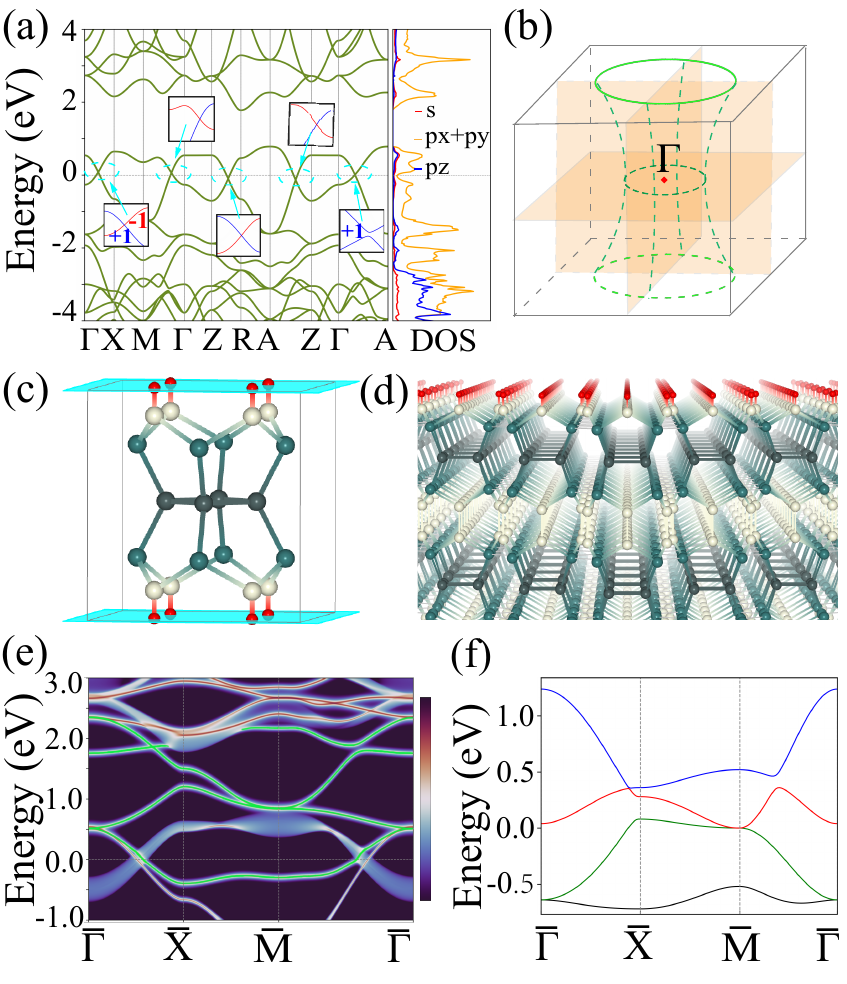}
\caption{
 (a) Projected band structure and density of states of bct-C$_{20}$, which is projected to $s$, $p_x+p_y$, and $p_z$ orbitals. The insets are the symmetry eigenvalues for mirror symmetries $m_{x,y,z}$, in which the red line denotes the eigenvalue of -1 and the blue line denotes the eigenvalue of +1. Thus the band crossings in $k_{x,y,z}=0,\pi$ plane are protected by mirror symmetries.
 (b) The schematic diagram of the nodal cage in the momentum space of bct-C$_{20}$.
 (c) The position of obstructed Wannier charge center of bct-C$_{20}$ that is colored in red.
 (d) The semi-infinite (001) surface of bct-C$_{20}$ with the OWCCs exposed on the surface. In this plot, only the OWCCs on the surface are shown.
 (e) Four FSSs highlighted in blue within the energy windows [-1~eV,2.5~eV].
 (f) The band structure of the effective surface Hamiltonian. The four floating bands $FS_{k=1,2,3,4}$ are plotted in black, green, red, and blue, respectively.
 }
\label{fig3}
\end{figure}

The calculated electronic bands of bct-C$_{20}$ along the high-symmetry lines in BZ are shown in  Fig.~\ref{fig3}(a).
The projected bands and density of states predict that the states near the Fermi level are dominated by the $E_u$, \textit{i.e.}, {$p_x+p_y$} orbitals of C atoms. Moreover, it is obvious that there are many band crossing points along the high symmetry lines, \textit{e.g.}, $\Gamma$-X, $\Gamma$-M, and Z-R.
By carefully checking, we find that the band crossings in the whole BZ form a nodal cage, as shown in Fig.~\ref{fig3}(b).
We detailed discussed the features, \textit{e.g.} the shape of the cage and the band topology of the nodal cage phase in SM~\cite{SM}.
Here, we show the projected local density of states for the (001) surface in Fig.~\ref{fig3}(e). As shown in Fig.~\ref{fig3}(e), there are many surface states with floating characteristics in the projected local density of states of the (001) surface, which we will discuss in the next section.

\textit{The abundant surface Dirac cones for OAI phase. ---}
Here, we prove that the FSSs in (001) surface derive from the nontrivial topological phase, \textit{i.e.}, obstructed atomic insulator phase of the gap in the energy windows $[1~\mathrm{eV}, 2~\mathrm{eV}]$. The number of occupied bands, \textit{i.e.}, $N_{Occ}$  corresponding to the energy gap is 41.
More importantly, for the first time, we show that the OAI phase provides an opportunity to achieve surface Dirac cones with different dispersion and band topologies.
Specifically, in bct-C$_{20}$, we found that due to the gap being in the OAI phase, there are two kinds of symmetry-restricted surface Dirac cones: topological Dirac cones with linear dispersion and symmetry-enforced quadratic Dirac cones.

To verify the band topology of the gap, here, we recall the definition of RSI. Based on the band representation, RSI is defined at specific Wyckoff position $W$. Non-zero RSI at $W$ indicates that there are $orbitals$  pinned at $W$. For the WP $4j$ whose site group is $C_{2v}$, highlighted by red balls in Fig.~\ref{fig3}(d), the $\mathbb{Z}$ type RSI is written as
\begin{equation}
\begin{split}
\delta \left(j\right)&=\sum_{i=1,2}m\left(B_{i}\right)-m\left(A_{i}\right)=-2,\\
\end{split}
\end{equation}
where the integer $m\left(\rho_{\omega_{k}}\right)$ means the multiplicity of the irreducible representation $\rho_{\omega_{k}}$ that is defined at WP 4$j$.
The RSI $\delta \left(j\right)=-2$ imply that there are at
least two $orbitals$, \textit{e.g.}, two $s$ $orbitals$ that are pinned at
the unoccupied $4j$ Wyckoff position.
Thus, the gap with $N_{Occ}=41$ is in the OAI phase.
As shown in Figs.~\ref{fig3}(d) and (e), some FSSs occur when the OWCCs of the system are exposed on the (001) surface.
It is worth noting that, the band connection of the FSSs is restricted by the symmetries that are maintained in the semi-infinite system with the open boundary condition applied for the [001] direction, as shown in Fig.~\ref{fig3}(d).
One observes in Fig.~\ref{fig3}(e) that there are four visible floating surface bands within the energy windows [-1~eV,2.5~eV].
These four bands are labeled as $FS_{k=1,2,3,4}$  ordered by energy at $\overline{\mathrm{X}}$ and highlighted in different colors, \textit{i.e.}, black, green, red, and blue in Fig.~\ref{fig3}(f).
There are two kinds of band crossings formed by FSSs at $\overline{\mathrm{M}}$, $\overline{\Gamma}$, and the high symmetry line $\overline{\Delta}$, \textit{i.e.}, $\overline{\mathrm{\Gamma}}\overline{\mathrm{X}}$.
Here, we analyze the band dispersion and band topology of the two kinds of band crossings.
The symmetries preserved in the semi-infinite system form a subgroup of $P4/mmm$, \textit{i.e.}, $P4mm$ (No.99).
And, the OWCC is exactly at the $4d$ $(x,x,z)$ WP of $P4mm$.
Based on the non-zero RSI, we put $s$ orbitals at the $4d$ WPs of $P4mm$ and construct an effective surface tight-binding Hamiltonian (see the derivation and the explicit expression of this Hamiltonian in SM~\cite{SM}).
With appropriate couplings considered, as shown in Fig.~\ref{fig3}(f) and SM~\cite{SM}, the band structure of the effective surface Hamiltonian keeps in good agreement with that obtained by the Wannier function and shown in Fig.~\ref{fig3}(e).

\begin{figure}[t]
\includegraphics[width=\linewidth]
        {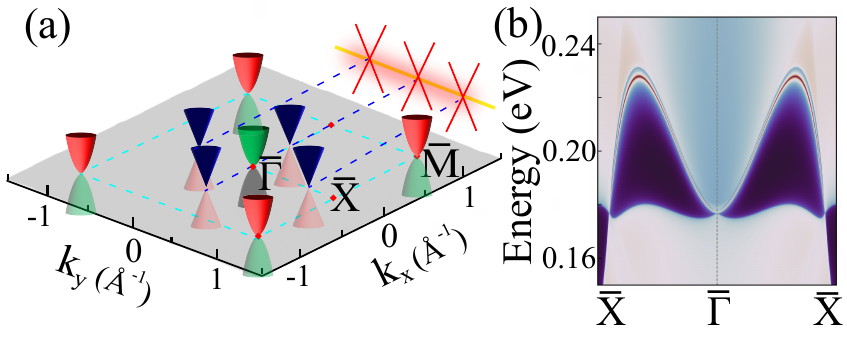}
	\caption{
 (a) The schematic diagram of the abundant surface Dirac cones, \textit{i.e.}, quadratic Dirac cones at $\overline{\Gamma}$ and $\overline{\mathrm{M}}$ and linear Dirac cones at $\overline{\Delta}$. The different colors of the Dirac cones represent different FSSs corresponding to that in Fig.~\ref{fig3}(f). For a 1D nano-ribbon, there will be hinge states (highlighted in orange) connecting the projects of linear surface Dirac cones (denoted by the red crosses).
 (b) The edge states of the effective surface Hamiltonian, which indicates the existence of hinge state in bct-C$_{20}$.
}
\label{fig4}
\end{figure}

Here, We provide a comprehensive symmetry analysis and low-energy effective models to gain a thorough understanding of the presence of the doubly degenerated point located at  $\overline{\mathrm{M}}$.
The generating elements of the space group $P4mm$ are  4-fold rotation symmetry $C_{4z}^+$, mirror symmetry $m_y$, and time-reversal symmetry. Significantly, the little group at $\overline{\mathrm{M}}$ 
has a 2D single-valued irreducible representation.
By the calculation of the eigenvalues of symmetries, we get that the little group representation of the band crossing formed by $FS_2$ and $FS_3$ is $\mathrm{M}_5$. Here, the band representation is written in BCS convention~\cite{liu2021spacegroupirep}.
Thus, this band degeneracy is symmetry enforced.
To build an effective model, the basis state is chosen
as $\psi=\left\{ \left|m=1\right\rangle,\left|m=-1\right\rangle \right\}$ with $m$ is the eigenvalue of $m_y$.
Under this basis, the matrix representation of $m_y$ and $\mathcal{T}$ read
\begin{equation}\label{HM}
\begin{array}{cc}
D\left(m_{y}\right)=\sigma_{z}, & D\left(\mathcal{T}\right)=-\sigma_{0}\mathcal{K},
\end{array}
\end{equation}
where $\sigma_{0}$ is the $2\times2$ identity matrix, $\sigma_{i} (i = x, y, z)$ is the Pauli
matrix, and $\mathcal{K}$ is the complex conjugate operator. Further, required by the following algebra
\begin{equation}
\begin{array}{ccc}
m_{x\left(y\right)}^{2}=1, & \left[m_{x},m_{y}\right]=0, & m_{x}m_{y}=\left(C_{4}^{+}\right)^{2}=-1\end{array},
\end{equation}
one can chose  $D\left(C_{4z}^+\right)=i \sigma_{y}$.
The effective Hamiltonian $H_{\overline{\mathrm{M}}}\left(\boldsymbol{k}\right)$ with this 2D band representation  is required to satisfy $D\left(\mathcal{O}\right)H_{\overline{\mathrm{M}}}\left(\mathcal{O}^{-1}\boldsymbol{k}\right)D^{-1}\left(\mathcal{O}\right)=H_{\overline{\mathrm{M}}}\left(\boldsymbol{k}\right)$. Hence, yielded by this symmetries, the explicit form of the minimal effective model is
\begin{equation}\label{HM}
\begin{split}
H_{\overline{\mathrm{M}}}\left(\boldsymbol{k}\right)&=c_1 (k_x^2-k_y^2)\sigma_z+c_2 k_x k_y \sigma_x .\\
\end{split}
\end{equation}
The absence of the linear function of $k_{x,y}$ in Eq.~(\ref{HM}) indicates that the dispersion of band crossing at $\overline{\mathrm{M}}$ is quadratic.
Similarly, for the band crossing formed by $FS_1$ and $FS_2$ at $\overline{\Gamma}$, since the little group representation is $\Gamma_5$, the crossing is also described by Eq.~(\ref{HM}), which indicates the quadratic band dispersion and same symmetry protection mechanism with that at $\overline{\mathrm{M}}$.
When we turn to the band crossing on $\overline{\Delta}$ which is formed by $FS_3$ and $FS_4$, the obtained little group representations of $FS_3$ and $FS_4$ at an arbitrary $k$ point are $\Delta_1$ and $\Delta_2$.
Due to  $\Delta_1$ and $\Delta_2$ carry the opposite eigenvalues of mirror symmetry $m_{100}$, the band crossing on $\overline{\Delta}$ is mirror protected and exhibits linear dispersion, as schematically shown in Fig.~\ref{fig4}(a).
It is known that the mirror-protected linear band crossings in 2D system possess nontrivial band topology. When clipped into 1D nano-ribbon, characteristic edge state connecting projected points is in expectation, as schematically shown in Fig.~\ref{fig4}(a).
To verify this point, the energy spectrum of 1D bct-C$_{20}$ with the periodicity along $x$-direction maintained is obtained and shown in Fig.~\ref{fig4}(b).
One observes that the emerging edge states indicate the system possesses metallic hinge states in bct-C$_{20}$.

\textit{Discussion and Summary. ---}
In summary, we established a connection between the realization of 2D quasi-particles and the 3D OAI. Based on first-principles calculations, we realized our proposal in a funnel-like carbon phase bct-C$_{20}$, which possesses the structure, energetic, dynamic, and thermal stabilities.
The calculated electronic structure and the symmetry analysis suggest that bct-C$_{20}$ possesses a nodal cage near the Fermi level. 
More importantly, we found a huge band gap with OAI band topology in the conduction band.
When the OWCCs are exposed on the (001) surface, there will be four in-gap FSSs.
The dispersion and band topology of the surface states are restricted by the symmetries maintained on the surface.
As a result, we get abundant surface Dirac cones, \textit{i.e.}, the symmetry-protected quadratic surface Dirac cones at $\overline{\Gamma}$ and $\overline{\mathrm{M}}$ and topological linear surface Dirac cones on high symmetry line $\overline{\Delta}$.
The band topology of the linear surface Dirac cone is verified by the hinge state.

We admit that the OAI gap in bct-$C_{20}$ is in the conduction band which would be an obstacle to the relevant experimental verification.
Fortunately, the new progress obtained in the field of OAIs or unconventional materials makes the intrinsic OAI with 2D quasi-particles in FSSs accessible~\cite{PhysRevB.105.165135, GAO2022598}. 
On the other hand, the emergence of the main interesting results of our work is beneficial from the featured FSSs. 
Naturally, the FSSs can form nodal line states. In the bulk states of 2D or 3D system, the nodal line could be protected by combination of inversion and time-reversal symmetry $\mathcal{PT}$, mirror symmetry $\mathcal{M}$, or non-symmorphic symmetry $\mathcal{\hat{O}}$. As for a surface of 3D system, both $\mathcal{P}$ and $\mathcal{M}$ parallel to the surface are destroyed. Hence, there would only be $\mathcal{\hat{O}}$ protected nodal line in the surface of 3D OAI.
Our work would absolutely inspire interest in research concerning revealing the 2D topological states, \textit{e.g.}, quantum spin hall effect, nodal line state, and topological flat bands in the FSSs of a higher-dimensional crystal.

\textit{Acknowledgement. ---}
This work was supported by the National Natural Science Foundation of China (NSFC, Grants No.~12204074, No.~12222402, No.~11974062, and No. 12147102).
D.-S. Ma also acknowledges the funding from the China National Postdoctoral Program for Innovative Talent (Grant No. BX20220367).


\begin{thebibliography}{79}%
	\makeatletter
	\providecommand \@ifxundefined [1]{%
	 \@ifx{#1\undefined}
	}%
	\providecommand \@ifnum [1]{%
	 \ifnum #1\expandafter \@firstoftwo
	 \else \expandafter \@secondoftwo
	 \fi
	}%
	\providecommand \@ifx [1]{%
	 \ifx #1\expandafter \@firstoftwo
	 \else \expandafter \@secondoftwo
	 \fi
	}%
	\providecommand \natexlab [1]{#1}%
	\providecommand \enquote  [1]{``#1''}%
	\providecommand \bibnamefont  [1]{#1}%
	\providecommand \bibfnamefont [1]{#1}%
	\providecommand \citenamefont [1]{#1}%
	\providecommand \href@noop [0]{\@secondoftwo}%
	\providecommand \href [0]{\begingroup \@sanitize@url \@href}%
	\providecommand \@href[1]{\@@startlink{#1}\@@href}%
	\providecommand \@@href[1]{\endgroup#1\@@endlink}%
	\providecommand \@sanitize@url [0]{\catcode `\\12\catcode `\$12\catcode
		`\&12\catcode `\#12\catcode `\^12\catcode `\_12\catcode `\%12\relax}%
	\providecommand \@@startlink[1]{}%
	\providecommand \@@endlink[0]{}%
	\providecommand \url  [0]{\begingroup\@sanitize@url \@url }%
	\providecommand \@url [1]{\endgroup\@href {#1}{\urlprefix }}%
	\providecommand \urlprefix  [0]{URL }%
	\providecommand \Eprint [0]{\href }%
	\providecommand \doibase [0]{http://dx.doi.org/}%
	\providecommand \selectlanguage [0]{\@gobble}%
	\providecommand \bibinfo  [0]{\@secondoftwo}%
	\providecommand \bibfield  [0]{\@secondoftwo}%
	\providecommand \translation [1]{[#1]}%
	\providecommand \BibitemOpen [0]{}%
	\providecommand \bibitemStop [0]{}%
	\providecommand \bibitemNoStop [0]{.\EOS\space}%
	\providecommand \EOS [0]{\spacefactor3000\relax}%
	\providecommand \BibitemShut  [1]{\csname bibitem#1\endcsname}%
	\let\auto@bib@innerbib\@empty
	\bibitem [{\citenamefont {Neto}\ \emph {et~al.}(2009)\citenamefont {Neto},
		\citenamefont {Guinea}, \citenamefont {Peres}, \citenamefont {Novoselov},\
		and\ \citenamefont {Geim}}]{neto2009electronic}%
		\BibitemOpen
		\bibfield  {author} {\bibinfo {author} {\bibfnamefont {A.~C.}\ \bibnamefont
		{Neto}}, \bibinfo {author} {\bibfnamefont {F.}~\bibnamefont {Guinea}},
		\bibinfo {author} {\bibfnamefont {N.~M.}\ \bibnamefont {Peres}}, \bibinfo
		{author} {\bibfnamefont {K.~S.}\ \bibnamefont {Novoselov}}, \ and\ \bibinfo
		{author} {\bibfnamefont {A.~K.}\ \bibnamefont {Geim}},\ }\href
		{https://link.aps.org/doi/10.1103/RevModPhys.81.109} {\bibfield  {journal}
		{\bibinfo  {journal} {Rev. Mod. Phys.}\ }\textbf {\bibinfo {volume} {81}},\
		\bibinfo {pages} {109} (\bibinfo {year} {2009})}\BibitemShut {NoStop}%
	\bibitem [{\citenamefont {Palumbo}\ and\ \citenamefont
		{Meichanetzidis}(2015)}]{palumbo2015two}%
		\BibitemOpen
		\bibfield  {author} {\bibinfo {author} {\bibfnamefont {G.}~\bibnamefont
		{Palumbo}}\ and\ \bibinfo {author} {\bibfnamefont {K.}~\bibnamefont
		{Meichanetzidis}},\ }\href
		{https://link.aps.org/doi/10.1103/PhysRevB.92.235106} {\bibfield  {journal}
		{\bibinfo  {journal} {Phys. Rev. B}\ }\textbf {\bibinfo {volume} {92}},\
		\bibinfo {pages} {235106} (\bibinfo {year} {2015})}\BibitemShut {NoStop}%
	\bibitem [{\citenamefont {Young}\ and\ \citenamefont
		{Kane}(2015)}]{young2015dirac}%
		\BibitemOpen
		\bibfield  {author} {\bibinfo {author} {\bibfnamefont {S.~M.}\ \bibnamefont
		{Young}}\ and\ \bibinfo {author} {\bibfnamefont {C.~L.}\ \bibnamefont
		{Kane}},\ }\href {https://link.aps.org/doi/10.1103/PhysRevLett.115.126803}
		{\bibfield  {journal} {\bibinfo  {journal} {Phys. Rev. Lett.}\ }\textbf
		{\bibinfo {volume} {115}},\ \bibinfo {pages} {126803} (\bibinfo {year}
		{2015})}\BibitemShut {NoStop}%
	\bibitem [{\citenamefont {Wang}\ \emph {et~al.}(2015)\citenamefont {Wang},
		\citenamefont {Ki}, \citenamefont {Chen}, \citenamefont {Berger},
		\citenamefont {MacDonald},\ and\ \citenamefont {Morpurgo}}]{wang2015strong}%
		\BibitemOpen
		\bibfield  {author} {\bibinfo {author} {\bibfnamefont {Z.}~\bibnamefont
		{Wang}}, \bibinfo {author} {\bibfnamefont {D.-K.}\ \bibnamefont {Ki}},
		\bibinfo {author} {\bibfnamefont {H.}~\bibnamefont {Chen}}, \bibinfo {author}
		{\bibfnamefont {H.}~\bibnamefont {Berger}}, \bibinfo {author} {\bibfnamefont
		{A.~H.}\ \bibnamefont {MacDonald}}, \ and\ \bibinfo {author} {\bibfnamefont
		{A.~F.}\ \bibnamefont {Morpurgo}},\ }\href
		{https://doi.org/10.1038/ncomms9339} {\bibfield  {journal} {\bibinfo
		{journal} {Nat. Commun.}\ }\textbf {\bibinfo {volume} {6}},\ \bibinfo {pages}
		{8339} (\bibinfo {year} {2015})}\BibitemShut {NoStop}%
	\bibitem [{\citenamefont {Feng}\ \emph {et~al.}(2017)\citenamefont {Feng},
		\citenamefont {Fu}, \citenamefont {Kasamatsu}, \citenamefont {Ito},
		\citenamefont {Cheng}, \citenamefont {Liu}, \citenamefont {Feng},
		\citenamefont {Wu}, \citenamefont {Mahatha}, \citenamefont {Sheverdyaeva}
		\emph {et~al.}}]{feng2017experimental}%
		\BibitemOpen
		\bibfield  {author} {\bibinfo {author} {\bibfnamefont {B.}~\bibnamefont
		{Feng}}, \bibinfo {author} {\bibfnamefont {B.}~\bibnamefont {Fu}}, \bibinfo
		{author} {\bibfnamefont {S.}~\bibnamefont {Kasamatsu}}, \bibinfo {author}
		{\bibfnamefont {S.}~\bibnamefont {Ito}}, \bibinfo {author} {\bibfnamefont
		{P.}~\bibnamefont {Cheng}}, \bibinfo {author} {\bibfnamefont {C.-C.}\
		\bibnamefont {Liu}}, \bibinfo {author} {\bibfnamefont {Y.}~\bibnamefont
		{Feng}}, \bibinfo {author} {\bibfnamefont {S.}~\bibnamefont {Wu}}, \bibinfo
		{author} {\bibfnamefont {S.~K.}\ \bibnamefont {Mahatha}}, \bibinfo {author}
		{\bibfnamefont {P.}~\bibnamefont {Sheverdyaeva}},  \emph {et~al.},\ }\href
		{https://doi.org/10.1038/s41467-017-01108-z} {\bibfield  {journal} {\bibinfo
		{journal} {Nat. Commun.}\ }\textbf {\bibinfo {volume} {8}},\ \bibinfo {pages}
		{1007} (\bibinfo {year} {2017})}\BibitemShut {NoStop}%
	\bibitem [{\citenamefont {Gao}\ \emph {et~al.}(2018)\citenamefont {Gao},
		\citenamefont {Sun}, \citenamefont {Lu}, \citenamefont {Li}, \citenamefont
		{Qian}, \citenamefont {Zhang}, \citenamefont {Zhang}, \citenamefont {Qian},
		\citenamefont {Ding}, \citenamefont {Lin} \emph {et~al.}}]{gao2018epitaxial}%
		\BibitemOpen
		\bibfield  {author} {\bibinfo {author} {\bibfnamefont {L.}~\bibnamefont
		{Gao}}, \bibinfo {author} {\bibfnamefont {J.-T.}\ \bibnamefont {Sun}},
		\bibinfo {author} {\bibfnamefont {J.-C.}\ \bibnamefont {Lu}}, \bibinfo
		{author} {\bibfnamefont {H.}~\bibnamefont {Li}}, \bibinfo {author}
		{\bibfnamefont {K.}~\bibnamefont {Qian}}, \bibinfo {author} {\bibfnamefont
		{S.}~\bibnamefont {Zhang}}, \bibinfo {author} {\bibfnamefont {Y.-Y.}\
		\bibnamefont {Zhang}}, \bibinfo {author} {\bibfnamefont {T.}~\bibnamefont
		{Qian}}, \bibinfo {author} {\bibfnamefont {H.}~\bibnamefont {Ding}}, \bibinfo
		{author} {\bibfnamefont {X.}~\bibnamefont {Lin}},  \emph {et~al.},\ }\href
		{https://onlinelibrary.wiley.com/doi/abs/10.1002/adma.201707055} {\bibfield
		{journal} {\bibinfo  {journal} {Adv. Mater.}\ }\textbf {\bibinfo {volume}
		{30}},\ \bibinfo {pages} {1707055} (\bibinfo {year} {2018})}\BibitemShut
		{NoStop}%
	\bibitem [{\citenamefont {Jeon}\ \emph {et~al.}(2019)\citenamefont {Jeon},
		\citenamefont {Oh},\ and\ \citenamefont {Kim}}]{jeon2019ferromagnetic}%
		\BibitemOpen
		\bibfield  {author} {\bibinfo {author} {\bibfnamefont {S.}~\bibnamefont
		{Jeon}}, \bibinfo {author} {\bibfnamefont {Y.-T.}\ \bibnamefont {Oh}}, \ and\
		\bibinfo {author} {\bibfnamefont {Y.}~\bibnamefont {Kim}},\ }\href
		{https://link.aps.org/doi/10.1103/PhysRevB.100.035406} {\bibfield  {journal}
		{\bibinfo  {journal} {Phys. Rev. B}\ }\textbf {\bibinfo {volume} {100}},\
		\bibinfo {pages} {035406} (\bibinfo {year} {2019})}\BibitemShut {NoStop}%
	\bibitem [{\citenamefont {Feng}\ \emph {et~al.}(2019)\citenamefont {Feng},
		\citenamefont {Zhang}, \citenamefont {Feng}, \citenamefont {Fu},
		\citenamefont {Wu}, \citenamefont {Miyamoto}, \citenamefont {He},
		\citenamefont {Chen}, \citenamefont {Wu}, \citenamefont {Shimada} \emph
		{et~al.}}]{feng2019discovery}%
		\BibitemOpen
		\bibfield  {author} {\bibinfo {author} {\bibfnamefont {B.}~\bibnamefont
		{Feng}}, \bibinfo {author} {\bibfnamefont {R.-W.}\ \bibnamefont {Zhang}},
		\bibinfo {author} {\bibfnamefont {Y.}~\bibnamefont {Feng}}, \bibinfo {author}
		{\bibfnamefont {B.}~\bibnamefont {Fu}}, \bibinfo {author} {\bibfnamefont
		{S.}~\bibnamefont {Wu}}, \bibinfo {author} {\bibfnamefont {K.}~\bibnamefont
		{Miyamoto}}, \bibinfo {author} {\bibfnamefont {S.}~\bibnamefont {He}},
		\bibinfo {author} {\bibfnamefont {L.}~\bibnamefont {Chen}}, \bibinfo {author}
		{\bibfnamefont {K.}~\bibnamefont {Wu}}, \bibinfo {author} {\bibfnamefont
		{K.}~\bibnamefont {Shimada}},  \emph {et~al.},\ }\href
		{https://link.aps.org/doi/10.1103/PhysRevLett.123.116401} {\bibfield
		{journal} {\bibinfo  {journal} {Phys. Rev. Lett.}\ }\textbf {\bibinfo
		{volume} {123}},\ \bibinfo {pages} {116401} (\bibinfo {year}
		{2019})}\BibitemShut {NoStop}%
	\bibitem [{\citenamefont {Niu}\ \emph {et~al.}(2019)\citenamefont {Niu},
		\citenamefont {Hanke}, \citenamefont {Buhl}, \citenamefont {Zhang},
		\citenamefont {Plucinski}, \citenamefont {Wortmann}, \citenamefont
		{Bl{\"u}gel}, \citenamefont {Bihlmayer},\ and\ \citenamefont
		{Mokrousov}}]{niu2019mixed}%
		\BibitemOpen
		\bibfield  {author} {\bibinfo {author} {\bibfnamefont {C.}~\bibnamefont
		{Niu}}, \bibinfo {author} {\bibfnamefont {J.-P.}\ \bibnamefont {Hanke}},
		\bibinfo {author} {\bibfnamefont {P.~M.}\ \bibnamefont {Buhl}}, \bibinfo
		{author} {\bibfnamefont {H.}~\bibnamefont {Zhang}}, \bibinfo {author}
		{\bibfnamefont {L.}~\bibnamefont {Plucinski}}, \bibinfo {author}
		{\bibfnamefont {D.}~\bibnamefont {Wortmann}}, \bibinfo {author}
		{\bibfnamefont {S.}~\bibnamefont {Bl{\"u}gel}}, \bibinfo {author}
		{\bibfnamefont {G.}~\bibnamefont {Bihlmayer}}, \ and\ \bibinfo {author}
		{\bibfnamefont {Y.}~\bibnamefont {Mokrousov}},\ }\href
		{https://doi.org/10.1038/s41467-019-10930-6} {\bibfield  {journal} {\bibinfo
		{journal} {Nat. Commun.}\ }\textbf {\bibinfo {volume} {10}},\ \bibinfo
		{pages} {3179} (\bibinfo {year} {2019})}\BibitemShut {NoStop}%
	\bibitem [{\citenamefont {Jin}\ \emph {et~al.}(2020)\citenamefont {Jin},
		\citenamefont {Zheng}, \citenamefont {Xiao}, \citenamefont {Chen},
		\citenamefont {Xu},\ and\ \citenamefont {Xu}}]{jin2020two}%
		\BibitemOpen
		\bibfield  {author} {\bibinfo {author} {\bibfnamefont {Y.~J.}\ \bibnamefont
		{Jin}}, \bibinfo {author} {\bibfnamefont {B.~B.}\ \bibnamefont {Zheng}},
		\bibinfo {author} {\bibfnamefont {X.~L.}\ \bibnamefont {Xiao}}, \bibinfo
		{author} {\bibfnamefont {Z.~J.}\ \bibnamefont {Chen}}, \bibinfo {author}
		{\bibfnamefont {Y.}~\bibnamefont {Xu}}, \ and\ \bibinfo {author}
		{\bibfnamefont {H.}~\bibnamefont {Xu}},\ }\href
		{https://link.aps.org/doi/10.1103/PhysRevLett.125.116402} {\bibfield
		{journal} {\bibinfo  {journal} {Phys. Rev. Lett.}\ }\textbf {\bibinfo
		{volume} {125}},\ \bibinfo {pages} {116402} (\bibinfo {year}
		{2020})}\BibitemShut {NoStop}%
	\bibitem [{\citenamefont {Wan}\ \emph {et~al.}(2011)\citenamefont {Wan},
		\citenamefont {Turner}, \citenamefont {Vishwanath},\ and\ \citenamefont
		{Savrasov}}]{wan2011topological}%
		\BibitemOpen
		\bibfield  {author} {\bibinfo {author} {\bibfnamefont {X.}~\bibnamefont
		{Wan}}, \bibinfo {author} {\bibfnamefont {A.~M.}\ \bibnamefont {Turner}},
		\bibinfo {author} {\bibfnamefont {A.}~\bibnamefont {Vishwanath}}, \ and\
		\bibinfo {author} {\bibfnamefont {S.~Y.}\ \bibnamefont {Savrasov}},\ }\href
		{https://link.aps.org/doi/10.1103/PhysRevB.83.205101} {\bibfield  {journal}
		{\bibinfo  {journal} {Phys. Rev. B}\ }\textbf {\bibinfo {volume} {83}},\
		\bibinfo {pages} {205101} (\bibinfo {year} {2011})}\BibitemShut {NoStop}%
	\bibitem [{\citenamefont {Chiu}\ and\ \citenamefont
		{Schnyder}(2014)}]{chiu2014classification}%
		\BibitemOpen
		\bibfield  {author} {\bibinfo {author} {\bibfnamefont {C.-K.}\ \bibnamefont
		{Chiu}}\ and\ \bibinfo {author} {\bibfnamefont {A.~P.}\ \bibnamefont
		{Schnyder}},\ }\href {https://link.aps.org/doi/10.1103/PhysRevB.90.205136}
		{\bibfield  {journal} {\bibinfo  {journal} {Phys. Rev. B}\ }\textbf {\bibinfo
		{volume} {90}},\ \bibinfo {pages} {205136} (\bibinfo {year}
		{2014})}\BibitemShut {NoStop}%
	\bibitem [{\citenamefont {Bansil}\ \emph {et~al.}(2016)\citenamefont {Bansil},
		\citenamefont {Lin},\ and\ \citenamefont {Das}}]{bansil2016colloquium}%
		\BibitemOpen
		\bibfield  {author} {\bibinfo {author} {\bibfnamefont {A.}~\bibnamefont
		{Bansil}}, \bibinfo {author} {\bibfnamefont {H.}~\bibnamefont {Lin}}, \ and\
		\bibinfo {author} {\bibfnamefont {T.}~\bibnamefont {Das}},\ }\href
		{https://link.aps.org/doi/10.1103/RevModPhys.88.021004} {\bibfield  {journal}
		{\bibinfo  {journal} {Rev. Mod. Phys.}\ }\textbf {\bibinfo {volume} {88}},\
		\bibinfo {pages} {021004} (\bibinfo {year} {2016})}\BibitemShut {NoStop}%
	\bibitem [{\citenamefont {Burkov}(2016)}]{burkov2016topological}%
		\BibitemOpen
		\bibfield  {author} {\bibinfo {author} {\bibfnamefont {A.}~\bibnamefont
		{Burkov}},\ }\href {https://doi.org/10.1038/nmat4788} {\bibfield  {journal}
		{\bibinfo  {journal} {Nat. Mater.}\ }\textbf {\bibinfo {volume} {15}},\
		\bibinfo {pages} {1145} (\bibinfo {year} {2016})}\BibitemShut {NoStop}%
	\bibitem [{\citenamefont {Luo}\ \emph {et~al.}(2020)\citenamefont {Luo},
		\citenamefont {Ji}, \citenamefont {Lu}, \citenamefont {Zhang},\ and\
		\citenamefont {Xiang}}]{PhysRevB.101.195111}%
		\BibitemOpen
		\bibfield  {author} {\bibinfo {author} {\bibfnamefont {W.}~\bibnamefont
		{Luo}}, \bibinfo {author} {\bibfnamefont {J.}~\bibnamefont {Ji}}, \bibinfo
		{author} {\bibfnamefont {J.}~\bibnamefont {Lu}}, \bibinfo {author}
		{\bibfnamefont {X.}~\bibnamefont {Zhang}}, \ and\ \bibinfo {author}
		{\bibfnamefont {H.}~\bibnamefont {Xiang}},\ }\href
		{https://link.aps.org/doi/10.1103/PhysRevB.101.195111} {\bibfield  {journal}
		{\bibinfo  {journal} {Phys. Rev. B}\ }\textbf {\bibinfo {volume} {101}},\
		\bibinfo {pages} {195111} (\bibinfo {year} {2020})}\BibitemShut {NoStop}%
	\bibitem [{\citenamefont {Xiao}\ \emph {et~al.}(2010)\citenamefont {Xiao},
		\citenamefont {Chang},\ and\ \citenamefont {Niu}}]{xiao2010berry}%
		\BibitemOpen
		\bibfield  {author} {\bibinfo {author} {\bibfnamefont {D.}~\bibnamefont
		{Xiao}}, \bibinfo {author} {\bibfnamefont {M.-C.}\ \bibnamefont {Chang}}, \
		and\ \bibinfo {author} {\bibfnamefont {Q.}~\bibnamefont {Niu}},\ }\href
		{https://link.aps.org/doi/10.1103/RevModPhys.82.1959} {\bibfield  {journal}
		{\bibinfo  {journal} {Rev. Mod. Phys.}\ }\textbf {\bibinfo {volume} {82}},\
		\bibinfo {pages} {1959} (\bibinfo {year} {2010})}\BibitemShut {NoStop}%
	\bibitem [{\citenamefont {Hosur}\ and\ \citenamefont
		{Qi}(2013)}]{hosur2013recent}%
		\BibitemOpen
		\bibfield  {author} {\bibinfo {author} {\bibfnamefont {P.}~\bibnamefont
		{Hosur}}\ and\ \bibinfo {author} {\bibfnamefont {X.}~\bibnamefont {Qi}},\
		}\href {https://www.sciencedirect.com/science/article/pii/S1631070513001710}
		{\bibfield  {journal} {\bibinfo  {journal} {C. R. Phys.}\ }\textbf {\bibinfo
		{volume} {14}},\ \bibinfo {pages} {857} (\bibinfo {year} {2013})}\BibitemShut
		{NoStop}%
	\bibitem [{\citenamefont {Wang}\ \emph {et~al.}(2016)\citenamefont {Wang},
		\citenamefont {Li}, \citenamefont {Liu}, \citenamefont {Yan}, \citenamefont
		{Wang}, \citenamefont {Liu}, \citenamefont {Lin}, \citenamefont {Li},
		\citenamefont {Wang}, \citenamefont {Li}, \citenamefont {Mandrus},
		\citenamefont {Xie}, \citenamefont {Feng},\ and\ \citenamefont
		{Wang}}]{PhysRevB.93.165127}%
		\BibitemOpen
		\bibfield  {author} {\bibinfo {author} {\bibfnamefont {H.}~\bibnamefont
		{Wang}}, \bibinfo {author} {\bibfnamefont {C.-K.}\ \bibnamefont {Li}},
		\bibinfo {author} {\bibfnamefont {H.}~\bibnamefont {Liu}}, \bibinfo {author}
		{\bibfnamefont {J.}~\bibnamefont {Yan}}, \bibinfo {author} {\bibfnamefont
		{J.}~\bibnamefont {Wang}}, \bibinfo {author} {\bibfnamefont {J.}~\bibnamefont
		{Liu}}, \bibinfo {author} {\bibfnamefont {Z.}~\bibnamefont {Lin}}, \bibinfo
		{author} {\bibfnamefont {Y.}~\bibnamefont {Li}}, \bibinfo {author}
		{\bibfnamefont {Y.}~\bibnamefont {Wang}}, \bibinfo {author} {\bibfnamefont
		{L.}~\bibnamefont {Li}}, \bibinfo {author} {\bibfnamefont {D.}~\bibnamefont
		{Mandrus}}, \bibinfo {author} {\bibfnamefont {X.~C.}\ \bibnamefont {Xie}},
		\bibinfo {author} {\bibfnamefont {J.}~\bibnamefont {Feng}}, \ and\ \bibinfo
		{author} {\bibfnamefont {J.}~\bibnamefont {Wang}},\ }\href
		{https://link.aps.org/doi/10.1103/PhysRevB.93.165127} {\bibfield  {journal}
		{\bibinfo  {journal} {Phys. Rev. B}\ }\textbf {\bibinfo {volume} {93}},\
		\bibinfo {pages} {165127} (\bibinfo {year} {2016})}\BibitemShut {NoStop}%
	\bibitem [{\citenamefont {Hu}\ \emph {et~al.}(2019)\citenamefont {Hu},
		\citenamefont {Xu}, \citenamefont {Ni},\ and\ \citenamefont
		{Mao}}]{hu2019transport}%
		\BibitemOpen
		\bibfield  {author} {\bibinfo {author} {\bibfnamefont {J.}~\bibnamefont
		{Hu}}, \bibinfo {author} {\bibfnamefont {S.-Y.}\ \bibnamefont {Xu}}, \bibinfo
		{author} {\bibfnamefont {N.}~\bibnamefont {Ni}}, \ and\ \bibinfo {author}
		{\bibfnamefont {Z.}~\bibnamefont {Mao}},\ }\href
		{https://doi.org/10.1146/annurev-matsci-070218-010023} {\bibfield  {journal}
		{\bibinfo  {journal} {Annu. Rev. Mater. Res.}\ }\textbf {\bibinfo {volume}
		{49}},\ \bibinfo {pages} {207} (\bibinfo {year} {2019})}\BibitemShut
		{NoStop}%
	\bibitem [{\citenamefont {Novoselov}\ \emph {et~al.}(2005)\citenamefont
		{Novoselov}, \citenamefont {Geim}, \citenamefont {Morozov}, \citenamefont
		{Jiang}, \citenamefont {Katsnelson}, \citenamefont {Grigorieva},
		\citenamefont {Dubonos},\ and\ \citenamefont {Firsov}}]{novoselov2005two}%
		\BibitemOpen
		\bibfield  {author} {\bibinfo {author} {\bibfnamefont {K.~S.}\ \bibnamefont
		{Novoselov}}, \bibinfo {author} {\bibfnamefont {A.~K.}\ \bibnamefont {Geim}},
		\bibinfo {author} {\bibfnamefont {S.~V.}\ \bibnamefont {Morozov}}, \bibinfo
		{author} {\bibfnamefont {D.}~\bibnamefont {Jiang}}, \bibinfo {author}
		{\bibfnamefont {M.~I.}\ \bibnamefont {Katsnelson}}, \bibinfo {author}
		{\bibfnamefont {I.~V.}\ \bibnamefont {Grigorieva}}, \bibinfo {author}
		{\bibfnamefont {S.}~\bibnamefont {Dubonos}}, \ and\ \bibinfo {author}
		{\bibfnamefont {a.}~\bibnamefont {Firsov}},\ }\href
		{https://doi.org/10.1038/nature04233} {\bibfield  {journal} {\bibinfo
		{journal} {Nature}\ }\textbf {\bibinfo {volume} {438}},\ \bibinfo {pages}
		{197} (\bibinfo {year} {2005})}\BibitemShut {NoStop}%
	\bibitem [{\citenamefont {Zhang}\ \emph {et~al.}(2005)\citenamefont {Zhang},
		\citenamefont {Tan}, \citenamefont {Stormer},\ and\ \citenamefont
		{Kim}}]{zhang2005experimental}%
		\BibitemOpen
		\bibfield  {author} {\bibinfo {author} {\bibfnamefont {Y.}~\bibnamefont
		{Zhang}}, \bibinfo {author} {\bibfnamefont {Y.-W.}\ \bibnamefont {Tan}},
		\bibinfo {author} {\bibfnamefont {H.~L.}\ \bibnamefont {Stormer}}, \ and\
		\bibinfo {author} {\bibfnamefont {P.}~\bibnamefont {Kim}},\ }\href
		{https://doi.org/10.1038/nature04235} {\bibfield  {journal} {\bibinfo
		{journal} {Nature}\ }\textbf {\bibinfo {volume} {438}},\ \bibinfo {pages}
		{201} (\bibinfo {year} {2005})}\BibitemShut {NoStop}%
	\bibitem [{\citenamefont {Novoselov}\ \emph
		{et~al.}(2007{\natexlab{a}})\citenamefont {Novoselov}, \citenamefont {Jiang},
		\citenamefont {Zhang}, \citenamefont {Morozov}, \citenamefont {Stormer},
		\citenamefont {Zeitler}, \citenamefont {Maan}, \citenamefont {Boebinger},
		\citenamefont {Kim},\ and\ \citenamefont {Geim}}]{novoselov2007room}%
		\BibitemOpen
		\bibfield  {author} {\bibinfo {author} {\bibfnamefont {K.~S.}\ \bibnamefont
		{Novoselov}}, \bibinfo {author} {\bibfnamefont {Z.}~\bibnamefont {Jiang}},
		\bibinfo {author} {\bibfnamefont {Y.}~\bibnamefont {Zhang}}, \bibinfo
		{author} {\bibfnamefont {S.}~\bibnamefont {Morozov}}, \bibinfo {author}
		{\bibfnamefont {H.~L.}\ \bibnamefont {Stormer}}, \bibinfo {author}
		{\bibfnamefont {U.}~\bibnamefont {Zeitler}}, \bibinfo {author} {\bibfnamefont
		{J.}~\bibnamefont {Maan}}, \bibinfo {author} {\bibfnamefont {G.}~\bibnamefont
		{Boebinger}}, \bibinfo {author} {\bibfnamefont {P.}~\bibnamefont {Kim}}, \
		and\ \bibinfo {author} {\bibfnamefont {A.~K.}\ \bibnamefont {Geim}},\ }\href
		{https://www.science.org/doi/abs/10.1126/science.1137201} {\bibfield
		{journal} {\bibinfo  {journal} {Science}\ }\textbf {\bibinfo {volume}
		{315}},\ \bibinfo {pages} {1379} (\bibinfo {year}
		{2007}{\natexlab{a}})}\BibitemShut {NoStop}%
	\bibitem [{\citenamefont {Zou}\ \emph {et~al.}(2022)\citenamefont {Zou},
		\citenamefont {Fu}, \citenamefont {Wang}, \citenamefont {Hu},\ and\
		\citenamefont {Shen}}]{zou2022half}%
		\BibitemOpen
		\bibfield  {author} {\bibinfo {author} {\bibfnamefont {J.-Y.}\ \bibnamefont
		{Zou}}, \bibinfo {author} {\bibfnamefont {B.}~\bibnamefont {Fu}}, \bibinfo
		{author} {\bibfnamefont {H.-W.}\ \bibnamefont {Wang}}, \bibinfo {author}
		{\bibfnamefont {Z.-A.}\ \bibnamefont {Hu}}, \ and\ \bibinfo {author}
		{\bibfnamefont {S.-Q.}\ \bibnamefont {Shen}},\ }\href
		{https://link.aps.org/doi/10.1103/PhysRevB.105.L201106} {\bibfield  {journal}
		{\bibinfo  {journal} {Phys. Rev. B}\ }\textbf {\bibinfo {volume} {105}},\
		\bibinfo {pages} {L201106} (\bibinfo {year} {2022})}\BibitemShut {NoStop}%
	\bibitem [{\citenamefont {Singh}\ \emph {et~al.}(2022)\citenamefont {Singh},
		\citenamefont {Kumar}, \citenamefont {Roychowdhury}, \citenamefont
		{Shekhar},\ and\ \citenamefont {Felser}}]{Singh_2022}%
		\BibitemOpen
		\bibfield  {author} {\bibinfo {author} {\bibfnamefont {S.}~\bibnamefont
		{Singh}}, \bibinfo {author} {\bibfnamefont {N.}~\bibnamefont {Kumar}},
		\bibinfo {author} {\bibfnamefont {S.}~\bibnamefont {Roychowdhury}}, \bibinfo
		{author} {\bibfnamefont {C.}~\bibnamefont {Shekhar}}, \ and\ \bibinfo
		{author} {\bibfnamefont {C.}~\bibnamefont {Felser}},\ }\href
		{https://dx.doi.org/10.1088/1361-648X/ac5d19} {\bibfield  {journal} {\bibinfo
		 {journal} {J. Phys. Condens. Matter}\ }\textbf {\bibinfo {volume} {34}},\
		\bibinfo {pages} {225802} (\bibinfo {year} {2022})}\BibitemShut {NoStop}%
	\bibitem [{\citenamefont {Fujiyama}\ \emph {et~al.}(2022)\citenamefont
		{Fujiyama}, \citenamefont {Maebashi}, \citenamefont {Tajima}, \citenamefont
		{Tsumuraya}, \citenamefont {Cui}, \citenamefont {Ogata},\ and\ \citenamefont
		{Kato}}]{fujiyama2022large}%
		\BibitemOpen
		\bibfield  {author} {\bibinfo {author} {\bibfnamefont {S.}~\bibnamefont
		{Fujiyama}}, \bibinfo {author} {\bibfnamefont {H.}~\bibnamefont {Maebashi}},
		\bibinfo {author} {\bibfnamefont {N.}~\bibnamefont {Tajima}}, \bibinfo
		{author} {\bibfnamefont {T.}~\bibnamefont {Tsumuraya}}, \bibinfo {author}
		{\bibfnamefont {H.~B.}\ \bibnamefont {Cui}}, \bibinfo {author} {\bibfnamefont
		{M.}~\bibnamefont {Ogata}}, \ and\ \bibinfo {author} {\bibfnamefont
		{R.}~\bibnamefont {Kato}},\ }\href
		{https://link.aps.org/doi/10.1103/PhysRevLett.128.027201} {\bibfield
		{journal} {\bibinfo  {journal} {Phys. Rev. Lett.}\ }\textbf {\bibinfo
		{volume} {128}},\ \bibinfo {pages} {027201} (\bibinfo {year}
		{2022})}\BibitemShut {NoStop}%
	\bibitem [{\citenamefont {CastroNeto}\ \emph {et~al.}(2009)\citenamefont
		{CastroNeto}, \citenamefont {Guinea}, \citenamefont {Peres}, \citenamefont
		{Novoselov},\ and\ \citenamefont {Geim}}]{RevModPhys.81.109}%
		\BibitemOpen
		\bibfield  {author} {\bibinfo {author} {\bibfnamefont {A.~H.}\ \bibnamefont
		{CastroNeto}}, \bibinfo {author} {\bibfnamefont {F.}~\bibnamefont {Guinea}},
		\bibinfo {author} {\bibfnamefont {N.~M.~R.}\ \bibnamefont {Peres}}, \bibinfo
		{author} {\bibfnamefont {K.~S.}\ \bibnamefont {Novoselov}}, \ and\ \bibinfo
		{author} {\bibfnamefont {A.~K.}\ \bibnamefont {Geim}},\ }\href
		{https://link.aps.org/doi/10.1103/RevModPhys.81.109} {\bibfield  {journal}
		{\bibinfo  {journal} {Rev. Mod. Phys.}\ }\textbf {\bibinfo {volume} {81}},\
		\bibinfo {pages} {109} (\bibinfo {year} {2009})}\BibitemShut {NoStop}%
	\bibitem [{\citenamefont {Ali}\ \emph {et~al.}(2014)\citenamefont {Ali},
		\citenamefont {Xiong}, \citenamefont {Flynn}, \citenamefont {Tao},
		\citenamefont {Gibson}, \citenamefont {Schoop}, \citenamefont {Liang},
		\citenamefont {Haldolaarachchige}, \citenamefont {Hirschberger},
		\citenamefont {Ong} \emph {et~al.}}]{ali2014large}%
		\BibitemOpen
		\bibfield  {author} {\bibinfo {author} {\bibfnamefont {M.~N.}\ \bibnamefont
		{Ali}}, \bibinfo {author} {\bibfnamefont {J.}~\bibnamefont {Xiong}}, \bibinfo
		{author} {\bibfnamefont {S.}~\bibnamefont {Flynn}}, \bibinfo {author}
		{\bibfnamefont {J.}~\bibnamefont {Tao}}, \bibinfo {author} {\bibfnamefont
		{Q.~D.}\ \bibnamefont {Gibson}}, \bibinfo {author} {\bibfnamefont {L.~M.}\
		\bibnamefont {Schoop}}, \bibinfo {author} {\bibfnamefont {T.}~\bibnamefont
		{Liang}}, \bibinfo {author} {\bibfnamefont {N.}~\bibnamefont
		{Haldolaarachchige}}, \bibinfo {author} {\bibfnamefont {M.}~\bibnamefont
		{Hirschberger}}, \bibinfo {author} {\bibfnamefont {N.~P.}\ \bibnamefont
		{Ong}},  \emph {et~al.},\ }\href {https://doi.org/10.1038/nature13763}
		{\bibfield  {journal} {\bibinfo  {journal} {Nature}\ }\textbf {\bibinfo
		{volume} {514}},\ \bibinfo {pages} {205} (\bibinfo {year}
		{2014})}\BibitemShut {NoStop}%
	\bibitem [{\citenamefont {Lai}\ \emph {et~al.}(2018)\citenamefont {Lai},
		\citenamefont {Liu}, \citenamefont {Ma}, \citenamefont {Wang}, \citenamefont
		{Zhang}, \citenamefont {Ren}, \citenamefont {Liu}, \citenamefont {Gu},
		\citenamefont {Zhuo}, \citenamefont {Lu} \emph
		{et~al.}}]{lai2018anisotropic}%
		\BibitemOpen
		\bibfield  {author} {\bibinfo {author} {\bibfnamefont {J.}~\bibnamefont
		{Lai}}, \bibinfo {author} {\bibfnamefont {X.}~\bibnamefont {Liu}}, \bibinfo
		{author} {\bibfnamefont {J.}~\bibnamefont {Ma}}, \bibinfo {author}
		{\bibfnamefont {Q.}~\bibnamefont {Wang}}, \bibinfo {author} {\bibfnamefont
		{K.}~\bibnamefont {Zhang}}, \bibinfo {author} {\bibfnamefont
		{X.}~\bibnamefont {Ren}}, \bibinfo {author} {\bibfnamefont {Y.}~\bibnamefont
		{Liu}}, \bibinfo {author} {\bibfnamefont {Q.}~\bibnamefont {Gu}}, \bibinfo
		{author} {\bibfnamefont {X.}~\bibnamefont {Zhuo}}, \bibinfo {author}
		{\bibfnamefont {W.}~\bibnamefont {Lu}},  \emph {et~al.},\ }\href
		{https://onlinelibrary.wiley.com/doi/abs/10.1002/adma.201707152} {\bibfield
		{journal} {\bibinfo  {journal} {Adv. Mater.}\ }\textbf {\bibinfo {volume}
		{30}},\ \bibinfo {pages} {1707152} (\bibinfo {year} {2018})}\BibitemShut
		{NoStop}%
	\bibitem [{\citenamefont {Wang}\ \emph
		{et~al.}(2019{\natexlab{a}})\citenamefont {Wang}, \citenamefont {Wang},
		\citenamefont {Liu}, \citenamefont {Wu}, \citenamefont {Wang}, \citenamefont
		{Yan}, \citenamefont {Cheng}, \citenamefont {Shi}, \citenamefont {Watanabe},
		\citenamefont {Taniguchi} \emph {et~al.}}]{wang2019direct}%
		\BibitemOpen
		\bibfield  {author} {\bibinfo {author} {\bibfnamefont {Y.}~\bibnamefont
		{Wang}}, \bibinfo {author} {\bibfnamefont {L.}~\bibnamefont {Wang}}, \bibinfo
		{author} {\bibfnamefont {X.}~\bibnamefont {Liu}}, \bibinfo {author}
		{\bibfnamefont {H.}~\bibnamefont {Wu}}, \bibinfo {author} {\bibfnamefont
		{P.}~\bibnamefont {Wang}}, \bibinfo {author} {\bibfnamefont {D.}~\bibnamefont
		{Yan}}, \bibinfo {author} {\bibfnamefont {B.}~\bibnamefont {Cheng}}, \bibinfo
		{author} {\bibfnamefont {Y.}~\bibnamefont {Shi}}, \bibinfo {author}
		{\bibfnamefont {K.}~\bibnamefont {Watanabe}}, \bibinfo {author}
		{\bibfnamefont {T.}~\bibnamefont {Taniguchi}},  \emph {et~al.},\ }\href
		{https://doi.org/10.1021/acs.nanolett.9b01275} {\bibfield  {journal}
		{\bibinfo  {journal} {Nano Lett.}\ }\textbf {\bibinfo {volume} {19}},\
		\bibinfo {pages} {3969} (\bibinfo {year} {2019}{\natexlab{a}})}\BibitemShut
		{NoStop}%
	\bibitem [{\citenamefont {Culcer}\ \emph {et~al.}(2020)\citenamefont {Culcer},
		\citenamefont {Keser}, \citenamefont {Li},\ and\ \citenamefont
		{Tkachov}}]{culcer2020transport}%
		\BibitemOpen
		\bibfield  {author} {\bibinfo {author} {\bibfnamefont {D.}~\bibnamefont
		{Culcer}}, \bibinfo {author} {\bibfnamefont {A.~C.}\ \bibnamefont {Keser}},
		\bibinfo {author} {\bibfnamefont {Y.}~\bibnamefont {Li}}, \ and\ \bibinfo
		{author} {\bibfnamefont {G.}~\bibnamefont {Tkachov}},\ }\href
		{https://dx.doi.org/10.1088/2053-1583/ab6ff7} {\bibfield  {journal} {\bibinfo
		 {journal} {2D Mater.}\ }\textbf {\bibinfo {volume} {7}},\ \bibinfo {pages}
		{022007} (\bibinfo {year} {2020})}\BibitemShut {NoStop}%
	\bibitem [{\citenamefont {Novoselov}\ \emph
		{et~al.}(2007{\natexlab{b}})\citenamefont {Novoselov}, \citenamefont {Jiang},
		\citenamefont {Zhang}, \citenamefont {Morozov}, \citenamefont {Stormer},
		\citenamefont {Zeitler}, \citenamefont {Maan}, \citenamefont {Boebinger},
		\citenamefont {Kim},\ and\ \citenamefont
		{Geim}}]{doi:10.1126/science.1137201}%
		\BibitemOpen
		\bibfield  {author} {\bibinfo {author} {\bibfnamefont {K.~S.}\ \bibnamefont
		{Novoselov}}, \bibinfo {author} {\bibfnamefont {Z.}~\bibnamefont {Jiang}},
		\bibinfo {author} {\bibfnamefont {Y.}~\bibnamefont {Zhang}}, \bibinfo
		{author} {\bibfnamefont {S.~V.}\ \bibnamefont {Morozov}}, \bibinfo {author}
		{\bibfnamefont {H.~L.}\ \bibnamefont {Stormer}}, \bibinfo {author}
		{\bibfnamefont {U.}~\bibnamefont {Zeitler}}, \bibinfo {author} {\bibfnamefont
		{J.~C.}\ \bibnamefont {Maan}}, \bibinfo {author} {\bibfnamefont {G.~S.}\
		\bibnamefont {Boebinger}}, \bibinfo {author} {\bibfnamefont {P.}~\bibnamefont
		{Kim}}, \ and\ \bibinfo {author} {\bibfnamefont {A.~K.}\ \bibnamefont
		{Geim}},\ }\href {https://www.science.org/doi/abs/10.1126/science.1137201}
		{\bibfield  {journal} {\bibinfo  {journal} {Science}\ }\textbf {\bibinfo
		{volume} {315}},\ \bibinfo {pages} {1379} (\bibinfo {year}
		{2007}{\natexlab{b}})}\BibitemShut {NoStop}%
	\bibitem [{\citenamefont {Ren}\ \emph {et~al.}(2016)\citenamefont {Ren},
		\citenamefont {Qiao},\ and\ \citenamefont {Niu}}]{ren2016topological}%
		\BibitemOpen
		\bibfield  {author} {\bibinfo {author} {\bibfnamefont {Y.}~\bibnamefont
		{Ren}}, \bibinfo {author} {\bibfnamefont {Z.}~\bibnamefont {Qiao}}, \ and\
		\bibinfo {author} {\bibfnamefont {Q.}~\bibnamefont {Niu}},\ }\href
		{https://dx.doi.org/10.1088/0034-4885/79/6/066501} {\bibfield  {journal}
		{\bibinfo  {journal} {Rep. Prog. Phys.}\ }\textbf {\bibinfo {volume} {79}},\
		\bibinfo {pages} {066501} (\bibinfo {year} {2016})}\BibitemShut {NoStop}%
	\bibitem [{\citenamefont {Wang}\ \emph
		{et~al.}(2019{\natexlab{b}})\citenamefont {Wang}, \citenamefont {Tang},
		\citenamefont {Ji}, \citenamefont {Zhang}, \citenamefont {Vishwanath},
		\citenamefont {Po},\ and\ \citenamefont {Wan}}]{wang2019two}%
		\BibitemOpen
		\bibfield  {author} {\bibinfo {author} {\bibfnamefont {D.}~\bibnamefont
		{Wang}}, \bibinfo {author} {\bibfnamefont {F.}~\bibnamefont {Tang}}, \bibinfo
		{author} {\bibfnamefont {J.}~\bibnamefont {Ji}}, \bibinfo {author}
		{\bibfnamefont {W.}~\bibnamefont {Zhang}}, \bibinfo {author} {\bibfnamefont
		{A.}~\bibnamefont {Vishwanath}}, \bibinfo {author} {\bibfnamefont {H.~C.}\
		\bibnamefont {Po}}, \ and\ \bibinfo {author} {\bibfnamefont {X.}~\bibnamefont
		{Wan}},\ }\href {https://link.aps.org/doi/10.1103/PhysRevB.100.195108}
		{\bibfield  {journal} {\bibinfo  {journal} {Phys. Rev. B}\ }\textbf {\bibinfo
		{volume} {100}},\ \bibinfo {pages} {195108} (\bibinfo {year}
		{2019}{\natexlab{b}})}\BibitemShut {NoStop}%
	\bibitem [{\citenamefont {Ma}\ \emph {et~al.}(2019)\citenamefont {Ma},
		\citenamefont {Deng}, \citenamefont {Zheng}, \citenamefont {Wu},
		\citenamefont {Liu}, \citenamefont {Zhou},\ and\ \citenamefont
		{Sun}}]{ma2019experimental}%
		\BibitemOpen
		\bibfield  {author} {\bibinfo {author} {\bibfnamefont {J.}~\bibnamefont
		{Ma}}, \bibinfo {author} {\bibfnamefont {K.}~\bibnamefont {Deng}}, \bibinfo
		{author} {\bibfnamefont {L.}~\bibnamefont {Zheng}}, \bibinfo {author}
		{\bibfnamefont {S.}~\bibnamefont {Wu}}, \bibinfo {author} {\bibfnamefont
		{Z.}~\bibnamefont {Liu}}, \bibinfo {author} {\bibfnamefont {S.}~\bibnamefont
		{Zhou}}, \ and\ \bibinfo {author} {\bibfnamefont {D.}~\bibnamefont {Sun}},\
		}\href {https://dx.doi.org/10.1088/2053-1583/ab0902} {\bibfield  {journal}
		{\bibinfo  {journal} {2D Mater.}\ }\textbf {\bibinfo {volume} {6}},\ \bibinfo
		{pages} {032001} (\bibinfo {year} {2019})}\BibitemShut {NoStop}%
	\bibitem [{\citenamefont {Bradlyn}\ \emph {et~al.}(2017)\citenamefont
		{Bradlyn}, \citenamefont {Elcoro}, \citenamefont {Cano}, \citenamefont
		{Vergniory}, \citenamefont {Wang}, \citenamefont {Felser}, \citenamefont
		{Aroyo},\ and\ \citenamefont {Bernevig}}]{bradlyn2017topological}%
		\BibitemOpen
		\bibfield  {author} {\bibinfo {author} {\bibfnamefont {B.}~\bibnamefont
		{Bradlyn}}, \bibinfo {author} {\bibfnamefont {L.}~\bibnamefont {Elcoro}},
		\bibinfo {author} {\bibfnamefont {J.}~\bibnamefont {Cano}}, \bibinfo {author}
		{\bibfnamefont {M.~G.}\ \bibnamefont {Vergniory}}, \bibinfo {author}
		{\bibfnamefont {Z.}~\bibnamefont {Wang}}, \bibinfo {author} {\bibfnamefont
		{C.}~\bibnamefont {Felser}}, \bibinfo {author} {\bibfnamefont {M.~I.}\
		\bibnamefont {Aroyo}}, \ and\ \bibinfo {author} {\bibfnamefont {B.~A.}\
		\bibnamefont {Bernevig}},\ }\href {https://doi.org/10.1038/nature23268}
		{\bibfield  {journal} {\bibinfo  {journal} {Nature}\ }\textbf {\bibinfo
		{volume} {547}},\ \bibinfo {pages} {298} (\bibinfo {year}
		{2017})}\BibitemShut {NoStop}%
	\bibitem [{\citenamefont {Elcoro}\ \emph {et~al.}(2021)\citenamefont {Elcoro},
		\citenamefont {Wieder}, \citenamefont {Song}, \citenamefont {Xu},
		\citenamefont {Bradlyn},\ and\ \citenamefont
		{Bernevig}}]{elcoro2021magnetic}%
		\BibitemOpen
		\bibfield  {author} {\bibinfo {author} {\bibfnamefont {L.}~\bibnamefont
		{Elcoro}}, \bibinfo {author} {\bibfnamefont {B.~J.}\ \bibnamefont {Wieder}},
		\bibinfo {author} {\bibfnamefont {Z.}~\bibnamefont {Song}}, \bibinfo {author}
		{\bibfnamefont {Y.}~\bibnamefont {Xu}}, \bibinfo {author} {\bibfnamefont
		{B.}~\bibnamefont {Bradlyn}}, \ and\ \bibinfo {author} {\bibfnamefont
		{B.~A.}\ \bibnamefont {Bernevig}},\ }\href
		{https://doi.org/10.1038/s41467-021-26241-8} {\bibfield  {journal} {\bibinfo
		{journal} {Nat. Commun.}\ }\textbf {\bibinfo {volume} {12}},\ \bibinfo
		{pages} {1} (\bibinfo {year} {2021})}\BibitemShut {NoStop}%
	\bibitem [{\citenamefont {Po}\ \emph {et~al.}(2017)\citenamefont {Po},
		\citenamefont {Vishwanath},\ and\ \citenamefont {Watanabe}}]{po2017symmetry}%
		\BibitemOpen
		\bibfield  {author} {\bibinfo {author} {\bibfnamefont {H.~C.}\ \bibnamefont
		{Po}}, \bibinfo {author} {\bibfnamefont {A.}~\bibnamefont {Vishwanath}}, \
		and\ \bibinfo {author} {\bibfnamefont {H.}~\bibnamefont {Watanabe}},\ }\href
		{https://doi.org/10.1038/s41467-017-00133-2} {\bibfield  {journal} {\bibinfo
		{journal} {Nat. Commun.}\ }\textbf {\bibinfo {volume} {8}},\ \bibinfo {pages}
		{1} (\bibinfo {year} {2017})}\BibitemShut {NoStop}%
	\bibitem [{\citenamefont {Kruthoff}\ \emph {et~al.}(2017)\citenamefont
		{Kruthoff}, \citenamefont {deBoer}, \citenamefont {vanWezel}, \citenamefont
		{Kane},\ and\ \citenamefont {Slager}}]{kruthoff2017topological}%
		\BibitemOpen
		\bibfield  {author} {\bibinfo {author} {\bibfnamefont {J.}~\bibnamefont
		{Kruthoff}}, \bibinfo {author} {\bibfnamefont {J.}~\bibnamefont {deBoer}},
		\bibinfo {author} {\bibfnamefont {J.}~\bibnamefont {vanWezel}}, \bibinfo
		{author} {\bibfnamefont {C.~L.}\ \bibnamefont {Kane}}, \ and\ \bibinfo
		{author} {\bibfnamefont {R.-J.}\ \bibnamefont {Slager}},\ }\href
		{https://link.aps.org/doi/10.1103/PhysRevX.7.041069} {\bibfield  {journal}
		{\bibinfo  {journal} {Phys. Rev. X}\ }\textbf {\bibinfo {volume} {7}},\
		\bibinfo {pages} {041069} (\bibinfo {year} {2017})}\BibitemShut {NoStop}%
	\bibitem [{\citenamefont {Watanabe}\ \emph {et~al.}(2018)\citenamefont
		{Watanabe}, \citenamefont {Po},\ and\ \citenamefont
		{Vishwanath}}]{watanabe2018structure}%
		\BibitemOpen
		\bibfield  {author} {\bibinfo {author} {\bibfnamefont {H.}~\bibnamefont
		{Watanabe}}, \bibinfo {author} {\bibfnamefont {H.~C.}\ \bibnamefont {Po}}, \
		and\ \bibinfo {author} {\bibfnamefont {A.}~\bibnamefont {Vishwanath}},\
		}\href {https://www.science.org/doi/abs/10.1126/sciadv.aat8685} {\bibfield
		{journal} {\bibinfo  {journal} {Sci. Adv.}\ }\textbf {\bibinfo {volume}
		{4}},\ \bibinfo {pages} {eaat8685} (\bibinfo {year} {2018})}\BibitemShut
		{NoStop}%
	\bibitem [{\citenamefont {Song}\ \emph {et~al.}(2018)\citenamefont {Song},
		\citenamefont {Zhang}, \citenamefont {Fang},\ and\ \citenamefont
		{Fang}}]{song2018quantitative}%
		\BibitemOpen
		\bibfield  {author} {\bibinfo {author} {\bibfnamefont {Z.}~\bibnamefont
		{Song}}, \bibinfo {author} {\bibfnamefont {T.}~\bibnamefont {Zhang}},
		\bibinfo {author} {\bibfnamefont {Z.}~\bibnamefont {Fang}}, \ and\ \bibinfo
		{author} {\bibfnamefont {C.}~\bibnamefont {Fang}},\ }\href
		{https://doi.org/10.1038/s41467-018-06010-w} {\bibfield  {journal} {\bibinfo
		{journal} {Nat. Commun.}\ }\textbf {\bibinfo {volume} {9}},\ \bibinfo {pages}
		{1} (\bibinfo {year} {2018})}\BibitemShut {NoStop}%
	\bibitem [{\citenamefont {Song}\ \emph
		{et~al.}(2020{\natexlab{a}})\citenamefont {Song}, \citenamefont {Elcoro},\
		and\ \citenamefont {Bernevig}}]{song2020twisted}%
		\BibitemOpen
		\bibfield  {author} {\bibinfo {author} {\bibfnamefont {Z.-D.}\ \bibnamefont
		{Song}}, \bibinfo {author} {\bibfnamefont {L.}~\bibnamefont {Elcoro}}, \ and\
		\bibinfo {author} {\bibfnamefont {B.~A.}\ \bibnamefont {Bernevig}},\ }\href
		{https://www.science.org/doi/abs/10.1126/science.aaz7650} {\bibfield
		{journal} {\bibinfo  {journal} {Science}\ }\textbf {\bibinfo {volume}
		{367}},\ \bibinfo {pages} {794} (\bibinfo {year}
		{2020}{\natexlab{a}})}\BibitemShut {NoStop}%
	\bibitem [{\citenamefont {Xu}\ \emph {et~al.}(2021{\natexlab{a}})\citenamefont
		{Xu}, \citenamefont {Elcoro}, \citenamefont {Li}, \citenamefont {Song},
		\citenamefont {Regnault}, \citenamefont {Yang}, \citenamefont {Sun},
		\citenamefont {Parkin}, \citenamefont {Felser},\ and\ \citenamefont
		{Bernevig}}]{xu2021three}%
		\BibitemOpen
		\bibfield  {author} {\bibinfo {author} {\bibfnamefont {Y.}~\bibnamefont
		{Xu}}, \bibinfo {author} {\bibfnamefont {L.}~\bibnamefont {Elcoro}}, \bibinfo
		{author} {\bibfnamefont {G.}~\bibnamefont {Li}}, \bibinfo {author}
		{\bibfnamefont {Z.-D.}\ \bibnamefont {Song}}, \bibinfo {author}
		{\bibfnamefont {N.}~\bibnamefont {Regnault}}, \bibinfo {author}
		{\bibfnamefont {Q.}~\bibnamefont {Yang}}, \bibinfo {author} {\bibfnamefont
		{Y.}~\bibnamefont {Sun}}, \bibinfo {author} {\bibfnamefont {S.}~\bibnamefont
		{Parkin}}, \bibinfo {author} {\bibfnamefont {C.}~\bibnamefont {Felser}}, \
		and\ \bibinfo {author} {\bibfnamefont {B.~A.}\ \bibnamefont {Bernevig}},\
		}\href {https://doi.org/10.48550/arXiv.2111.02433} {\bibfield  {journal}
		{\bibinfo  {journal} {arXiv preprint arXiv:2111.02433}\ } (\bibinfo {year}
		{2021}{\natexlab{a}})}\BibitemShut {NoStop}%
	\bibitem [{\citenamefont {Peng}\ \emph {et~al.}(2022)\citenamefont {Peng},
		\citenamefont {Jiang}, \citenamefont {Fang}, \citenamefont {Weng},\ and\
		\citenamefont {Fang}}]{peng2022topological}%
		\BibitemOpen
		\bibfield  {author} {\bibinfo {author} {\bibfnamefont {B.}~\bibnamefont
		{Peng}}, \bibinfo {author} {\bibfnamefont {Y.}~\bibnamefont {Jiang}},
		\bibinfo {author} {\bibfnamefont {Z.}~\bibnamefont {Fang}}, \bibinfo {author}
		{\bibfnamefont {H.}~\bibnamefont {Weng}}, \ and\ \bibinfo {author}
		{\bibfnamefont {C.}~\bibnamefont {Fang}},\ }\href
		{https://link.aps.org/doi/10.1103/PhysRevB.105.235138} {\bibfield  {journal}
		{\bibinfo  {journal} {Phys. Rev. B}\ }\textbf {\bibinfo {volume} {105}},\
		\bibinfo {pages} {235138} (\bibinfo {year} {2022})}\BibitemShut {NoStop}%
	\bibitem [{\citenamefont {Bouhon}\ \emph {et~al.}(2021)\citenamefont {Bouhon},
		\citenamefont {Lange},\ and\ \citenamefont {Slager}}]{bouhon2021topological}%
		\BibitemOpen
		\bibfield  {author} {\bibinfo {author} {\bibfnamefont {A.}~\bibnamefont
		{Bouhon}}, \bibinfo {author} {\bibfnamefont {G.~F.}\ \bibnamefont {Lange}}, \
		and\ \bibinfo {author} {\bibfnamefont {R.-J.}\ \bibnamefont {Slager}},\
		}\href {https://link.aps.org/doi/10.1103/PhysRevB.103.245127} {\bibfield
		{journal} {\bibinfo  {journal} {Phys. Rev. B}\ }\textbf {\bibinfo {volume}
		{103}},\ \bibinfo {pages} {245127} (\bibinfo {year} {2021})}\BibitemShut
		{NoStop}%
	\bibitem [{\citenamefont {Li}\ \emph {et~al.}(2022{\natexlab{a}})\citenamefont
		{Li}, \citenamefont {Ma}, \citenamefont {Liu}, \citenamefont {Yu},\ and\
		\citenamefont {Yao}}]{li2022atomic}%
		\BibitemOpen
		\bibfield  {author} {\bibinfo {author} {\bibfnamefont {X.-P.}\ \bibnamefont
		{Li}}, \bibinfo {author} {\bibfnamefont {D.-S.}\ \bibnamefont {Ma}}, \bibinfo
		{author} {\bibfnamefont {C.-C.}\ \bibnamefont {Liu}}, \bibinfo {author}
		{\bibfnamefont {Z.-M.}\ \bibnamefont {Yu}}, \ and\ \bibinfo {author}
		{\bibfnamefont {Y.}~\bibnamefont {Yao}},\ }\href
		{https://link.aps.org/doi/10.1103/PhysRevB.105.165135} {\bibfield  {journal}
		{\bibinfo  {journal} {Phys. Rev. B}\ }\textbf {\bibinfo {volume} {105}},\
		\bibinfo {pages} {165135} (\bibinfo {year} {2022}{\natexlab{a}})}\BibitemShut
		{NoStop}%
	\bibitem [{\citenamefont {Wang}\ \emph
		{et~al.}(2022{\natexlab{a}})\citenamefont {Wang}, \citenamefont {Jiang},
		\citenamefont {Liu}, \citenamefont {Zhang}, \citenamefont {Li}, \citenamefont
		{Liu}, \citenamefont {Sun}, \citenamefont {Weng},\ and\ \citenamefont
		{Chen}}]{wang2022two}%
		\BibitemOpen
		\bibfield  {author} {\bibinfo {author} {\bibfnamefont {L.}~\bibnamefont
		{Wang}}, \bibinfo {author} {\bibfnamefont {Y.}~\bibnamefont {Jiang}},
		\bibinfo {author} {\bibfnamefont {J.}~\bibnamefont {Liu}}, \bibinfo {author}
		{\bibfnamefont {S.}~\bibnamefont {Zhang}}, \bibinfo {author} {\bibfnamefont
		{J.}~\bibnamefont {Li}}, \bibinfo {author} {\bibfnamefont {P.}~\bibnamefont
		{Liu}}, \bibinfo {author} {\bibfnamefont {Y.}~\bibnamefont {Sun}}, \bibinfo
		{author} {\bibfnamefont {H.}~\bibnamefont {Weng}}, \ and\ \bibinfo {author}
		{\bibfnamefont {X.-Q.}\ \bibnamefont {Chen}},\ }\href
		{https://link.aps.org/doi/10.1103/PhysRevB.106.155144} {\bibfield  {journal}
		{\bibinfo  {journal} {Phys. Rev. B}\ }\textbf {\bibinfo {volume} {106}},\
		\bibinfo {pages} {155144} (\bibinfo {year} {2022}{\natexlab{a}})}\BibitemShut
		{NoStop}%
	\bibitem [{\citenamefont {Gao}\ \emph {et~al.}(2022{\natexlab{a}})\citenamefont
		{Gao}, \citenamefont {Qian}, \citenamefont {Jia}, \citenamefont {Guo},
		\citenamefont {Fang}, \citenamefont {Liu}, \citenamefont {Weng},\ and\
		\citenamefont {Wang}}]{gao2022unconventional}%
		\BibitemOpen
		\bibfield  {author} {\bibinfo {author} {\bibfnamefont {J.}~\bibnamefont
		{Gao}}, \bibinfo {author} {\bibfnamefont {Y.}~\bibnamefont {Qian}}, \bibinfo
		{author} {\bibfnamefont {H.}~\bibnamefont {Jia}}, \bibinfo {author}
		{\bibfnamefont {Z.}~\bibnamefont {Guo}}, \bibinfo {author} {\bibfnamefont
		{Z.}~\bibnamefont {Fang}}, \bibinfo {author} {\bibfnamefont {M.}~\bibnamefont
		{Liu}}, \bibinfo {author} {\bibfnamefont {H.}~\bibnamefont {Weng}}, \ and\
		\bibinfo {author} {\bibfnamefont {Z.}~\bibnamefont {Wang}},\ }\href
		{https://www.sciencedirect.com/science/article/pii/S2095927321008045}
		{\bibfield  {journal} {\bibinfo  {journal} {Sci. Bull.}\ }\textbf {\bibinfo
		{volume} {67}},\ \bibinfo {pages} {598} (\bibinfo {year}
		{2022}{\natexlab{a}})}\BibitemShut {NoStop}%
	\bibitem [{\citenamefont {Song}\ \emph
		{et~al.}(2020{\natexlab{b}})\citenamefont {Song}, \citenamefont {Elcoro},\
		and\ \citenamefont {Bernevig}}]{doi:10.1126/science.aaz7650}%
		\BibitemOpen
		\bibfield  {author} {\bibinfo {author} {\bibfnamefont {Z.-D.}\ \bibnamefont
		{Song}}, \bibinfo {author} {\bibfnamefont {L.}~\bibnamefont {Elcoro}}, \ and\
		\bibinfo {author} {\bibfnamefont {B.~A.}\ \bibnamefont {Bernevig}},\ }\href
		{https://www.science.org/doi/abs/10.1126/science.aaz7650} {\bibfield
		{journal} {\bibinfo  {journal} {Science}\ }\textbf {\bibinfo {volume}
		{367}},\ \bibinfo {pages} {794} (\bibinfo {year}
		{2020}{\natexlab{b}})}\BibitemShut {NoStop}%
	\bibitem [{\citenamefont {Xu}\ \emph {et~al.}(2021{\natexlab{b}})\citenamefont
		{Xu}, \citenamefont {Elcoro}, \citenamefont {Song}, \citenamefont
		{Vergniory}, \citenamefont {Felser}, \citenamefont {Parkin}, \citenamefont
		{Regnault}, \citenamefont {Ma{\~n}es},\ and\ \citenamefont
		{Bernevig}}]{xu2021filling}%
		\BibitemOpen
		\bibfield  {author} {\bibinfo {author} {\bibfnamefont {Y.}~\bibnamefont
		{Xu}}, \bibinfo {author} {\bibfnamefont {L.}~\bibnamefont {Elcoro}}, \bibinfo
		{author} {\bibfnamefont {Z.-D.}\ \bibnamefont {Song}}, \bibinfo {author}
		{\bibfnamefont {M.}~\bibnamefont {Vergniory}}, \bibinfo {author}
		{\bibfnamefont {C.}~\bibnamefont {Felser}}, \bibinfo {author} {\bibfnamefont
		{S.~S.}\ \bibnamefont {Parkin}}, \bibinfo {author} {\bibfnamefont
		{N.}~\bibnamefont {Regnault}}, \bibinfo {author} {\bibfnamefont {J.~L.}\
		\bibnamefont {Ma{\~n}es}}, \ and\ \bibinfo {author} {\bibfnamefont {B.~A.}\
		\bibnamefont {Bernevig}},\ }\href {https://doi.org/10.48550/arXiv.2106.10276}
		{\bibfield  {journal} {\bibinfo  {journal} {arXiv preprint arXiv:2106.10276}\
		} (\bibinfo {year} {2021}{\natexlab{b}})}\BibitemShut {NoStop}%
	\bibitem [{\citenamefont {Nie}\ \emph {et~al.}(2021)\citenamefont {Nie},
		\citenamefont {Qian}, \citenamefont {Gao}, \citenamefont {Fang},
		\citenamefont {Weng},\ and\ \citenamefont {Wang}}]{PhysRevB.103.205133}%
		\BibitemOpen
		\bibfield  {author} {\bibinfo {author} {\bibfnamefont {S.}~\bibnamefont
		{Nie}}, \bibinfo {author} {\bibfnamefont {Y.}~\bibnamefont {Qian}}, \bibinfo
		{author} {\bibfnamefont {J.}~\bibnamefont {Gao}}, \bibinfo {author}
		{\bibfnamefont {Z.}~\bibnamefont {Fang}}, \bibinfo {author} {\bibfnamefont
		{H.}~\bibnamefont {Weng}}, \ and\ \bibinfo {author} {\bibfnamefont
		{Z.}~\bibnamefont {Wang}},\ }\href
		{https://link.aps.org/doi/10.1103/PhysRevB.103.205133} {\bibfield  {journal}
		{\bibinfo  {journal} {Phys. Rev. B}\ }\textbf {\bibinfo {volume} {103}},\
		\bibinfo {pages} {205133} (\bibinfo {year} {2021})}\BibitemShut {NoStop}%
	\bibitem [{\citenamefont {Zhang}\ \emph {et~al.}(2022)\citenamefont {Zhang},
		\citenamefont {Deng}, \citenamefont {Sun}, \citenamefont {Fang},
		\citenamefont {Guo},\ and\ \citenamefont {Wang}}]{zhang2022large}%
		\BibitemOpen
		\bibfield  {author} {\bibinfo {author} {\bibfnamefont {R.}~\bibnamefont
		{Zhang}}, \bibinfo {author} {\bibfnamefont {J.}~\bibnamefont {Deng}},
		\bibinfo {author} {\bibfnamefont {Y.}~\bibnamefont {Sun}}, \bibinfo {author}
		{\bibfnamefont {Z.}~\bibnamefont {Fang}}, \bibinfo {author} {\bibfnamefont
		{Z.}~\bibnamefont {Guo}}, \ and\ \bibinfo {author} {\bibfnamefont
		{Z.}~\bibnamefont {Wang}},\ }\href
		{https://doi.org/10.48550/arXiv.2211.04116} {\bibfield  {journal} {\bibinfo
		{journal} {arXiv preprint arXiv:2211.04116}\ } (\bibinfo {year}
		{2022})}\BibitemShut {NoStop}%
	\bibitem [{\citenamefont {Schindler}\ \emph {et~al.}(2020)\citenamefont
		{Schindler}, \citenamefont {Bradlyn}, \citenamefont {Fischer},\ and\
		\citenamefont {Neupert}}]{PhysRevLett.124.247001}%
		\BibitemOpen
		\bibfield  {author} {\bibinfo {author} {\bibfnamefont {F.}~\bibnamefont
		{Schindler}}, \bibinfo {author} {\bibfnamefont {B.}~\bibnamefont {Bradlyn}},
		\bibinfo {author} {\bibfnamefont {M.~H.}\ \bibnamefont {Fischer}}, \ and\
		\bibinfo {author} {\bibfnamefont {T.}~\bibnamefont {Neupert}},\ }\href
		{https://link.aps.org/doi/10.1103/PhysRevLett.124.247001} {\bibfield
		{journal} {\bibinfo  {journal} {Phys. Rev. Lett.}\ }\textbf {\bibinfo
		{volume} {124}},\ \bibinfo {pages} {247001} (\bibinfo {year}
		{2020})}\BibitemShut {NoStop}%
	\bibitem [{\citenamefont {Aihara}\ \emph {et~al.}(2020)\citenamefont {Aihara},
		\citenamefont {Hirayama},\ and\ \citenamefont
		{Murakami}}]{PhysRevResearch.2.033224}%
		\BibitemOpen
		\bibfield  {author} {\bibinfo {author} {\bibfnamefont {Y.}~\bibnamefont
		{Aihara}}, \bibinfo {author} {\bibfnamefont {M.}~\bibnamefont {Hirayama}}, \
		and\ \bibinfo {author} {\bibfnamefont {S.}~\bibnamefont {Murakami}},\ }\href
		{\doibase 10.1103/PhysRevResearch.2.033224} {\bibfield  {journal} {\bibinfo
		{journal} {Phys. Rev. Res.}\ }\textbf {\bibinfo {volume} {2}},\ \bibinfo
		{pages} {033224} (\bibinfo {year} {2020})}\BibitemShut {NoStop}%
	\bibitem [{\citenamefont {Wang}\ \emph
		{et~al.}(2022{\natexlab{b}})\citenamefont {Wang}, \citenamefont {Jiang},
		\citenamefont {Liu}, \citenamefont {Zhang}, \citenamefont {Li}, \citenamefont
		{Liu}, \citenamefont {Sun}, \citenamefont {Weng},\ and\ \citenamefont
		{Chen}}]{PhysRevB.106.155144}%
		\BibitemOpen
		\bibfield  {author} {\bibinfo {author} {\bibfnamefont {L.}~\bibnamefont
		{Wang}}, \bibinfo {author} {\bibfnamefont {Y.}~\bibnamefont {Jiang}},
		\bibinfo {author} {\bibfnamefont {J.}~\bibnamefont {Liu}}, \bibinfo {author}
		{\bibfnamefont {S.}~\bibnamefont {Zhang}}, \bibinfo {author} {\bibfnamefont
		{J.}~\bibnamefont {Li}}, \bibinfo {author} {\bibfnamefont {P.}~\bibnamefont
		{Liu}}, \bibinfo {author} {\bibfnamefont {Y.}~\bibnamefont {Sun}}, \bibinfo
		{author} {\bibfnamefont {H.}~\bibnamefont {Weng}}, \ and\ \bibinfo {author}
		{\bibfnamefont {X.-Q.}\ \bibnamefont {Chen}},\ }\href
		{https://link.aps.org/doi/10.1103/PhysRevB.106.155144} {\bibfield  {journal}
		{\bibinfo  {journal} {Phys. Rev. B}\ }\textbf {\bibinfo {volume} {106}},\
		\bibinfo {pages} {155144} (\bibinfo {year} {2022}{\natexlab{b}})}\BibitemShut
		{NoStop}%
	\bibitem [{\citenamefont {Li}\ \emph {et~al.}(2022{\natexlab{b}})\citenamefont
		{Li}, \citenamefont {Ma}, \citenamefont {Liu}, \citenamefont {Yu},\ and\
		\citenamefont {Yao}}]{PhysRevB.105.165135}%
		\BibitemOpen
		\bibfield  {author} {\bibinfo {author} {\bibfnamefont {X.-P.}\ \bibnamefont
		{Li}}, \bibinfo {author} {\bibfnamefont {D.-S.}\ \bibnamefont {Ma}}, \bibinfo
		{author} {\bibfnamefont {C.-C.}\ \bibnamefont {Liu}}, \bibinfo {author}
		{\bibfnamefont {Z.-M.}\ \bibnamefont {Yu}}, \ and\ \bibinfo {author}
		{\bibfnamefont {Y.}~\bibnamefont {Yao}},\ }\href
		{https://link.aps.org/doi/10.1103/PhysRevB.105.165135} {\bibfield  {journal}
		{\bibinfo  {journal} {Phys. Rev. B}\ }\textbf {\bibinfo {volume} {105}},\
		\bibinfo {pages} {165135} (\bibinfo {year} {2022}{\natexlab{b}})}\BibitemShut
		{NoStop}%
	\bibitem [{\citenamefont {Gao}\ \emph {et~al.}(2022{\natexlab{b}})\citenamefont
		{Gao}, \citenamefont {Qian}, \citenamefont {Jia}, \citenamefont {Guo},
		\citenamefont {Fang}, \citenamefont {Liu}, \citenamefont {Weng},\ and\
		\citenamefont {Wang}}]{GAO2022598}%
		\BibitemOpen
		\bibfield  {author} {\bibinfo {author} {\bibfnamefont {J.}~\bibnamefont
		{Gao}}, \bibinfo {author} {\bibfnamefont {Y.}~\bibnamefont {Qian}}, \bibinfo
		{author} {\bibfnamefont {H.}~\bibnamefont {Jia}}, \bibinfo {author}
		{\bibfnamefont {Z.}~\bibnamefont {Guo}}, \bibinfo {author} {\bibfnamefont
		{Z.}~\bibnamefont {Fang}}, \bibinfo {author} {\bibfnamefont {M.}~\bibnamefont
		{Liu}}, \bibinfo {author} {\bibfnamefont {H.}~\bibnamefont {Weng}}, \ and\
		\bibinfo {author} {\bibfnamefont {Z.}~\bibnamefont {Wang}},\ }\href
		{https://www.sciencedirect.com/science/article/pii/S2095927321008045}
		{\bibfield  {journal} {\bibinfo  {journal} {Sci. Bull.}\ }\textbf {\bibinfo
		{volume} {67}},\ \bibinfo {pages} {598} (\bibinfo {year}
		{2022}{\natexlab{b}})}\BibitemShut {NoStop}%
	\bibitem [{\citenamefont {Banerjee}\ \emph {et~al.}(2020)\citenamefont
		{Banerjee}, \citenamefont {Mandal},\ and\ \citenamefont
		{Liew}}]{PhysRevLett.124.063901}%
		\BibitemOpen
		\bibfield  {author} {\bibinfo {author} {\bibfnamefont {R.}~\bibnamefont
		{Banerjee}}, \bibinfo {author} {\bibfnamefont {S.}~\bibnamefont {Mandal}}, \
		and\ \bibinfo {author} {\bibfnamefont {T.~C.~H.}\ \bibnamefont {Liew}},\
		}\href {https://link.aps.org/doi/10.1103/PhysRevLett.124.063901} {\bibfield
		{journal} {\bibinfo  {journal} {Phys. Rev. Lett.}\ }\textbf {\bibinfo
		{volume} {124}},\ \bibinfo {pages} {063901} (\bibinfo {year}
		{2020})}\BibitemShut {NoStop}%
	\bibitem [{\citenamefont {Kim}\ \emph {et~al.}(2023)\citenamefont {Kim},
		\citenamefont {Huang}, \citenamefont {Lin}, \citenamefont {Vanderbilt},\ and\
		\citenamefont {Kioussis}}]{kim2023bismuth}%
		\BibitemOpen
		\bibfield  {author} {\bibinfo {author} {\bibfnamefont {J.}~\bibnamefont
		{Kim}}, \bibinfo {author} {\bibfnamefont {C.-Y.}\ \bibnamefont {Huang}},
		\bibinfo {author} {\bibfnamefont {H.}~\bibnamefont {Lin}}, \bibinfo {author}
		{\bibfnamefont {D.}~\bibnamefont {Vanderbilt}}, \ and\ \bibinfo {author}
		{\bibfnamefont {N.}~\bibnamefont {Kioussis}},\ }\href
		{https://doi.org/10.1103/PhysRevB.107.045135} {\bibfield  {journal} {\bibinfo
		 {journal} {arXiv preprint arXiv:2301.04278}\ } (\bibinfo {year}
		{2023})}\BibitemShut {NoStop}%
	\bibitem [{\citenamefont {Benalcazar}\ \emph {et~al.}(2019)\citenamefont
		{Benalcazar}, \citenamefont {Li},\ and\ \citenamefont
		{Hughes}}]{benalcazar2019quantization}%
		\BibitemOpen
		\bibfield  {author} {\bibinfo {author} {\bibfnamefont {W.~A.}\ \bibnamefont
		{Benalcazar}}, \bibinfo {author} {\bibfnamefont {T.}~\bibnamefont {Li}}, \
		and\ \bibinfo {author} {\bibfnamefont {T.~L.}\ \bibnamefont {Hughes}},\
		}\href {https://link.aps.org/doi/10.1103/PhysRevB.99.245151} {\bibfield
		{journal} {\bibinfo  {journal} {Phys. Rev. B}\ }\textbf {\bibinfo {volume}
		{99}},\ \bibinfo {pages} {245151} (\bibinfo {year} {2019})}\BibitemShut
		{NoStop}%
	\bibitem [{SM()}]{SM}%
		\BibitemOpen
		\href@noop {} {\bibinfo  {journal} {See supplemental material for details of
		the calculations, the additional discussion of the stability of bct-C$_{20}$,
		the topological nodal cage in the bulk state, and the effective surface model
		Hamiltonian which include
		Refs.~\cite{PhysRevB.54.11169,PhysRev.136.B864,PhysRev.140.A1133,PhysRevLett.77.3865,PhysRevB.13.5188,doi:10.1063/1.1564060,RevModPhys.73.515,TOGO20151,RevModPhys.84.1419,
		MOSTOFI2008685,GAO2021107760,occelli2003properties,PhysRevB.90.224104,hill1952elastic,ma2018mirror,kim2023bismuth,
		song2020twisted, xu2021three}}\ }\BibitemShut {NoStop}%
	\bibitem [{\citenamefont {Sheng}\ \emph {et~al.}(2011)\citenamefont {Sheng},
		\citenamefont {Yan}, \citenamefont {Ye}, \citenamefont {Zheng},\ and\
		\citenamefont {Su}}]{T-carbon}%
		\BibitemOpen
	\bibfield  {journal} {  }\bibfield  {author} {\bibinfo {author} {\bibfnamefont
		{X.~L.}\ \bibnamefont {Sheng}}, \bibinfo {author} {\bibfnamefont {Q.~B.}\
		\bibnamefont {Yan}}, \bibinfo {author} {\bibfnamefont {F.}~\bibnamefont
		{Ye}}, \bibinfo {author} {\bibfnamefont {Q.~R.}\ \bibnamefont {Zheng}}, \
		and\ \bibinfo {author} {\bibfnamefont {G.}~\bibnamefont {Su}},\ }\href
		{https://link.aps.org/doi/10.1103/PhysRevLett.106.155703} {\bibfield
		{journal} {\bibinfo  {journal} {Phys. Rev. Lett.}\ }\textbf {\bibinfo
		{volume} {106}},\ \bibinfo {pages} {155703} (\bibinfo {year}
		{2011})}\BibitemShut {NoStop}%
	\bibitem [{\citenamefont {Ding}\ \emph {et~al.}(2020)\citenamefont {Ding},
		\citenamefont {Zhang}, \citenamefont {Gan}, \citenamefont {Cao},
		\citenamefont {Chen},\ and\ \citenamefont {Wang}}]{bct-C16}%
		\BibitemOpen
		\bibfield  {author} {\bibinfo {author} {\bibfnamefont {X.-Y.}\ \bibnamefont
		{Ding}}, \bibinfo {author} {\bibfnamefont {C.}~\bibnamefont {Zhang}},
		\bibinfo {author} {\bibfnamefont {L.-Y.}\ \bibnamefont {Gan}}, \bibinfo
		{author} {\bibfnamefont {Y.}~\bibnamefont {Cao}}, \bibinfo {author}
		{\bibfnamefont {L.-L.}\ \bibnamefont {Chen}}, \ and\ \bibinfo {author}
		{\bibfnamefont {R.}~\bibnamefont {Wang}},\ }\href
		{https://doi.org/10.1088/1367-2630/ab990b} {\bibfield  {journal} {\bibinfo
		{journal} {New J. Phys.}\ }\textbf {\bibinfo {volume} {22}},\ \bibinfo
		{pages} {073036} (\bibinfo {year} {2020})}\BibitemShut {NoStop}%
	\bibitem [{\citenamefont {Zhang}\ \emph {et~al.}(2020)\citenamefont {Zhang},
		\citenamefont {Ding}, \citenamefont {Gan}, \citenamefont {Cao}, \citenamefont
		{Li}, \citenamefont {Wu},\ and\ \citenamefont {Wang}}]{Fco-C6}%
		\BibitemOpen
		\bibfield  {author} {\bibinfo {author} {\bibfnamefont {C.}~\bibnamefont
		{Zhang}}, \bibinfo {author} {\bibfnamefont {X.-Y.}\ \bibnamefont {Ding}},
		\bibinfo {author} {\bibfnamefont {L.-Y.}\ \bibnamefont {Gan}}, \bibinfo
		{author} {\bibfnamefont {Y.}~\bibnamefont {Cao}}, \bibinfo {author}
		{\bibfnamefont {B.-S.}\ \bibnamefont {Li}}, \bibinfo {author} {\bibfnamefont
		{X.}~\bibnamefont {Wu}}, \ and\ \bibinfo {author} {\bibfnamefont
		{R.}~\bibnamefont {Wang}},\ }\href
		{https://link.aps.org/doi/10.1103/PhysRevB.101.235119} {\bibfield  {journal}
		{\bibinfo  {journal} {Phys. Rev. B}\ }\textbf {\bibinfo {volume} {101}},\
		\bibinfo {pages} {235119} (\bibinfo {year} {2020})}\BibitemShut {NoStop}%
	\bibitem [{\citenamefont {Mouhat}\ and\ \citenamefont
		{Coudert}(2014)}]{PhysRevB.90.224104}%
		\BibitemOpen
		\bibfield  {author} {\bibinfo {author} {\bibfnamefont {F.}~\bibnamefont
		{Mouhat}}\ and\ \bibinfo {author} {\bibfnamefont {F.~X.}\ \bibnamefont
		{Coudert}},\ }\href {https://link.aps.org/doi/10.1103/PhysRevB.90.224104}
		{\bibfield  {journal} {\bibinfo  {journal} {Phys. Rev. B}\ }\textbf {\bibinfo
		{volume} {90}},\ \bibinfo {pages} {224104} (\bibinfo {year}
		{2014})}\BibitemShut {NoStop}%
	\bibitem [{\citenamefont {Liu}\ \emph {et~al.}(2021)\citenamefont {Liu},
		\citenamefont {Chu}, \citenamefont {Zhang}, \citenamefont {Yu},\ and\
		\citenamefont {Yao}}]{liu2021spacegroupirep}%
		\BibitemOpen
		\bibfield  {author} {\bibinfo {author} {\bibfnamefont {G.-B.}\ \bibnamefont
		{Liu}}, \bibinfo {author} {\bibfnamefont {M.}~\bibnamefont {Chu}}, \bibinfo
		{author} {\bibfnamefont {Z.}~\bibnamefont {Zhang}}, \bibinfo {author}
		{\bibfnamefont {Z.-M.}\ \bibnamefont {Yu}}, \ and\ \bibinfo {author}
		{\bibfnamefont {Y.}~\bibnamefont {Yao}},\ }\href@noop {} {\bibfield
		{journal} {\bibinfo  {journal} {Comput. Phys. Commun.}\ }\textbf {\bibinfo
		{volume} {265}},\ \bibinfo {pages} {107993} (\bibinfo {year}
		{2021})}\BibitemShut {NoStop}%
	\bibitem [{\citenamefont {Kresse}\ and\ \citenamefont
		{Furthm\"uller}(1996)}]{PhysRevB.54.11169}%
		\BibitemOpen
		\bibfield  {author} {\bibinfo {author} {\bibfnamefont {G.}~\bibnamefont
		{Kresse}}\ and\ \bibinfo {author} {\bibfnamefont {J.}~\bibnamefont
		{Furthm\"uller}},\ }\href
		{https://link.aps.org/doi/10.1103/PhysRevB.54.11169} {\bibfield  {journal}
		{\bibinfo  {journal} {Phys. Rev. B}\ }\textbf {\bibinfo {volume} {54}},\
		\bibinfo {pages} {11169} (\bibinfo {year} {1996})}\BibitemShut {NoStop}%
	\bibitem [{\citenamefont {Hohenberg}\ and\ \citenamefont
		{Kohn}(1964)}]{PhysRev.136.B864}%
		\BibitemOpen
		\bibfield  {author} {\bibinfo {author} {\bibfnamefont {P.}~\bibnamefont
		{Hohenberg}}\ and\ \bibinfo {author} {\bibfnamefont {W.}~\bibnamefont
		{Kohn}},\ }\href {https://link.aps.org/doi/10.1103/PhysRev.136.B864}
		{\bibfield  {journal} {\bibinfo  {journal} {Phys. Rev.}\ }\textbf {\bibinfo
		{volume} {136}},\ \bibinfo {pages} {B864} (\bibinfo {year}
		{1964})}\BibitemShut {NoStop}%
	\bibitem [{\citenamefont {Kohn}\ and\ \citenamefont
		{Sham}(1965)}]{PhysRev.140.A1133}%
		\BibitemOpen
		\bibfield  {author} {\bibinfo {author} {\bibfnamefont {W.}~\bibnamefont
		{Kohn}}\ and\ \bibinfo {author} {\bibfnamefont {L.~J.}\ \bibnamefont
		{Sham}},\ }\href {https://link.aps.org/doi/10.1103/PhysRev.140.A1133}
		{\bibfield  {journal} {\bibinfo  {journal} {Phys. Rev.}\ }\textbf {\bibinfo
		{volume} {140}},\ \bibinfo {pages} {A1133} (\bibinfo {year}
		{1965})}\BibitemShut {NoStop}%
	\bibitem [{\citenamefont {Perdew}\ \emph {et~al.}(1996)\citenamefont {Perdew},
		\citenamefont {Burke},\ and\ \citenamefont
		{Ernzerhof}}]{PhysRevLett.77.3865}%
		\BibitemOpen
		\bibfield  {author} {\bibinfo {author} {\bibfnamefont {J.~P.}\ \bibnamefont
		{Perdew}}, \bibinfo {author} {\bibfnamefont {K.}~\bibnamefont {Burke}}, \
		and\ \bibinfo {author} {\bibfnamefont {M.}~\bibnamefont {Ernzerhof}},\ }\href
		{https://link.aps.org/doi/10.1103/PhysRevLett.77.3865} {\bibfield  {journal}
		{\bibinfo  {journal} {Phys. Rev. Lett.}\ }\textbf {\bibinfo {volume} {77}},\
		\bibinfo {pages} {3865} (\bibinfo {year} {1996})}\BibitemShut {NoStop}%
	\bibitem [{\citenamefont {Monkhorst}\ and\ \citenamefont
		{Pack}(1976)}]{PhysRevB.13.5188}%
		\BibitemOpen
		\bibfield  {author} {\bibinfo {author} {\bibfnamefont {H.~J.}\ \bibnamefont
		{Monkhorst}}\ and\ \bibinfo {author} {\bibfnamefont {J.~D.}\ \bibnamefont
		{Pack}},\ }\href {https://link.aps.org/doi/10.1103/PhysRevB.13.5188}
		{\bibfield  {journal} {\bibinfo  {journal} {Phys. Rev. B}\ }\textbf {\bibinfo
		{volume} {13}},\ \bibinfo {pages} {5188} (\bibinfo {year}
		{1976})}\BibitemShut {NoStop}%
	\bibitem [{\citenamefont {Heyd}\ \emph {et~al.}(2003)\citenamefont {Heyd},
		\citenamefont {Scuseria},\ and\ \citenamefont
		{Ernzerhof}}]{doi:10.1063/1.1564060}%
		\BibitemOpen
		\bibfield  {author} {\bibinfo {author} {\bibfnamefont {J.}~\bibnamefont
		{Heyd}}, \bibinfo {author} {\bibfnamefont {G.~E.}\ \bibnamefont {Scuseria}},
		\ and\ \bibinfo {author} {\bibfnamefont {M.}~\bibnamefont {Ernzerhof}},\
		}\href {https://doi.org/10.1063/1.1564060} {\bibfield  {journal} {\bibinfo
		{journal} {J. Chem. Phys.}\ }\textbf {\bibinfo {volume} {118}},\ \bibinfo
		{pages} {8207} (\bibinfo {year} {2003})}\BibitemShut {NoStop}%
	\bibitem [{\citenamefont {Baroni}\ \emph {et~al.}(2001)\citenamefont {Baroni},
		\citenamefont {de~Gironcoli}, \citenamefont {Dal~Corso},\ and\ \citenamefont
		{Giannozzi}}]{RevModPhys.73.515}%
		\BibitemOpen
		\bibfield  {author} {\bibinfo {author} {\bibfnamefont {S.}~\bibnamefont
		{Baroni}}, \bibinfo {author} {\bibfnamefont {S.}~\bibnamefont
		{de~Gironcoli}}, \bibinfo {author} {\bibfnamefont {A.}~\bibnamefont
		{Dal~Corso}}, \ and\ \bibinfo {author} {\bibfnamefont {P.}~\bibnamefont
		{Giannozzi}},\ }\href {https://link.aps.org/doi/10.1103/RevModPhys.73.515}
		{\bibfield  {journal} {\bibinfo  {journal} {Rev. Mod. Phys.}\ }\textbf
		{\bibinfo {volume} {73}},\ \bibinfo {pages} {515} (\bibinfo {year}
		{2001})}\BibitemShut {NoStop}%
	\bibitem [{\citenamefont {Togo}\ and\ \citenamefont
		{Tanaka}(2015)}]{TOGO20151}%
		\BibitemOpen
		\bibfield  {author} {\bibinfo {author} {\bibfnamefont {A.}~\bibnamefont
		{Togo}}\ and\ \bibinfo {author} {\bibfnamefont {I.}~\bibnamefont {Tanaka}},\
		}\href {https://www.sciencedirect.com/science/article/pii/S1359646215003127}
		{\bibfield  {journal} {\bibinfo  {journal} {Scr. Mater.}\ }\textbf {\bibinfo
		{volume} {108}},\ \bibinfo {pages} {1} (\bibinfo {year} {2015})}\BibitemShut
		{NoStop}%
	\bibitem [{\citenamefont {Marzari}\ \emph {et~al.}(2012)\citenamefont
		{Marzari}, \citenamefont {Mostofi}, \citenamefont {Yates}, \citenamefont
		{Souza},\ and\ \citenamefont {Vanderbilt}}]{RevModPhys.84.1419}%
		\BibitemOpen
		\bibfield  {author} {\bibinfo {author} {\bibfnamefont {N.}~\bibnamefont
		{Marzari}}, \bibinfo {author} {\bibfnamefont {A.~A.}\ \bibnamefont
		{Mostofi}}, \bibinfo {author} {\bibfnamefont {J.~R.}\ \bibnamefont {Yates}},
		\bibinfo {author} {\bibfnamefont {I.}~\bibnamefont {Souza}}, \ and\ \bibinfo
		{author} {\bibfnamefont {D.}~\bibnamefont {Vanderbilt}},\ }\href
		{https://link.aps.org/doi/10.1103/RevModPhys.84.1419} {\bibfield  {journal}
		{\bibinfo  {journal} {Rev. Mod. Phys.}\ }\textbf {\bibinfo {volume} {84}},\
		\bibinfo {pages} {1419} (\bibinfo {year} {2012})}\BibitemShut {NoStop}%
	\bibitem [{\citenamefont {Mostofi}\ \emph {et~al.}(2008)\citenamefont
		{Mostofi}, \citenamefont {Yates}, \citenamefont {Lee}, \citenamefont {Souza},
		\citenamefont {Vanderbilt},\ and\ \citenamefont {Marzari}}]{MOSTOFI2008685}%
		\BibitemOpen
		\bibfield  {author} {\bibinfo {author} {\bibfnamefont {A.~A.}\ \bibnamefont
		{Mostofi}}, \bibinfo {author} {\bibfnamefont {J.~R.}\ \bibnamefont {Yates}},
		\bibinfo {author} {\bibfnamefont {Y.-S.}\ \bibnamefont {Lee}}, \bibinfo
		{author} {\bibfnamefont {I.}~\bibnamefont {Souza}}, \bibinfo {author}
		{\bibfnamefont {D.}~\bibnamefont {Vanderbilt}}, \ and\ \bibinfo {author}
		{\bibfnamefont {N.}~\bibnamefont {Marzari}},\ }\href
		{https://www.sciencedirect.com/science/article/pii/S0010465507004936}
		{\bibfield  {journal} {\bibinfo  {journal} {Comput. Phys. Commun.}\ }\textbf
		{\bibinfo {volume} {178}},\ \bibinfo {pages} {685} (\bibinfo {year}
		{2008})}\BibitemShut {NoStop}%
	\bibitem [{\citenamefont {Gao}\ \emph {et~al.}(2021)\citenamefont {Gao},
		\citenamefont {Wu}, \citenamefont {Persson},\ and\ \citenamefont
		{Wang}}]{GAO2021107760}%
		\BibitemOpen
		\bibfield  {author} {\bibinfo {author} {\bibfnamefont {J.}~\bibnamefont
		{Gao}}, \bibinfo {author} {\bibfnamefont {Q.}~\bibnamefont {Wu}}, \bibinfo
		{author} {\bibfnamefont {C.}~\bibnamefont {Persson}}, \ and\ \bibinfo
		{author} {\bibfnamefont {Z.}~\bibnamefont {Wang}},\ }\href
		{https://www.sciencedirect.com/science/article/pii/S0010465520303805}
		{\bibfield  {journal} {\bibinfo  {journal} {Comput. Phys. Commun.}\ }\textbf
		{\bibinfo {volume} {261}},\ \bibinfo {pages} {107760} (\bibinfo {year}
		{2021})}\BibitemShut {NoStop}%
	\bibitem [{\citenamefont {Occelli}\ \emph {et~al.}(2003)\citenamefont
		{Occelli}, \citenamefont {Loubeyre},\ and\ \citenamefont
		{LeToullec}}]{occelli2003properties}%
		\BibitemOpen
		\bibfield  {author} {\bibinfo {author} {\bibfnamefont {F.}~\bibnamefont
		{Occelli}}, \bibinfo {author} {\bibfnamefont {P.}~\bibnamefont {Loubeyre}}, \
		and\ \bibinfo {author} {\bibfnamefont {R.}~\bibnamefont {LeToullec}},\ }\href
		{https://doi.org/10.1038/nmat831} {\bibfield  {journal} {\bibinfo  {journal}
		{Nat. Mater.}\ }\textbf {\bibinfo {volume} {2}},\ \bibinfo {pages} {151}
		(\bibinfo {year} {2003})}\BibitemShut {NoStop}%
	\bibitem [{\citenamefont {Hill}(1952)}]{hill1952elastic}%
		\BibitemOpen
		\bibfield  {author} {\bibinfo {author} {\bibfnamefont {R.}~\bibnamefont
		{Hill}},\ }\href {https://dx.doi.org/10.1088/0370-1298/65/5/307} {\bibfield
		{journal} {\bibinfo  {journal} {Proc. Phys. Soc. London}\ }\textbf {\bibinfo
		{volume} {65}},\ \bibinfo {pages} {349} (\bibinfo {year} {1952})}\BibitemShut
		{NoStop}%
	\bibitem [{\citenamefont {Ma}\ \emph {et~al.}(2018)\citenamefont {Ma},
		\citenamefont {Zhou}, \citenamefont {Fu}, \citenamefont {Yu}, \citenamefont
		{Liu},\ and\ \citenamefont {Yao}}]{ma2018mirror}%
		\BibitemOpen
		\bibfield  {author} {\bibinfo {author} {\bibfnamefont {D.-S.}\ \bibnamefont
		{Ma}}, \bibinfo {author} {\bibfnamefont {J.}~\bibnamefont {Zhou}}, \bibinfo
		{author} {\bibfnamefont {B.}~\bibnamefont {Fu}}, \bibinfo {author}
		{\bibfnamefont {Z.-M.}\ \bibnamefont {Yu}}, \bibinfo {author} {\bibfnamefont
		{C.-C.}\ \bibnamefont {Liu}}, \ and\ \bibinfo {author} {\bibfnamefont
		{Y.}~\bibnamefont {Yao}},\ }\href
		{https://link.aps.org/doi/10.1103/PhysRevB.98.201104} {\bibfield  {journal}
		{\bibinfo  {journal} {Phys. Rev. B}\ }\textbf {\bibinfo {volume} {98}},\
		\bibinfo {pages} {201104(R)} (\bibinfo {year} {2018})}\BibitemShut {NoStop}%
	\end{thebibliography}
\end{document}